\documentclass[12pt]{article}

\usepackage{amsfonts,amsmath,amssymb}
\usepackage{amssymb}
\usepackage{epsfig}
\usepackage[rflt]{floatflt}

\begin{document}

\def\p{\phi}\def\P{\Phi}\def\a{\alpha}\def\e{\epsilon}\def\be{\begin{equation}}
\def\ee{\end{equation}}\def\l{\label}\def\0{\setcounter{equation}{0}}
\def\b{\beta}\def\S{\Sigma}\def\C{\cite}
\def\r{\ref}\def\ba{\begin{eqnarray}}\def\ea{\end{eqnarray}}
\def\n{\nonumber}\def\R{\rho}\def\X{\Xi}\def\x{\xi}\def\la{\lambda}
\def\d{\delta}\def\s{\sigma}\def\f{\frac}\def\D{\Delta}\def\pa{\partial}
\def\Th{\Theta}\def\o{\omega}\def\O{\Omega}\def\th{\theta}\def\ga{\gamma}
\def\Ga{\Gamma}\def\t{\times}\def\h{\hat}\def\rar{\rightarrow}
\def\vp{\varphi}\def\inf{\infty}\def\le{\left}\def\ri{\right}
\def\foot{\footnote}\def\vep{\varepsilon}\def\N{\bar{n}(s)}
\def\k{\kappa}\def\sq{\sqrt{s}}\def\bx{{\mathbf x}}\def\La{\Lambda}
\def\bb{{\bf b}}\def\bq{{\bf q}}\def\cp{{\cal P}}\def\tg{\tilde{g}}
\def\cf{{\cal F}}\def\bN{{\bf N}}\def\Re{{\rm Re}}\def\Im{{\rm Im}}
\def\bk{\hat{\mathbf{k}}}\def\cl{{\cal L}}\def\cs{{\cal S}}\def\cn{{\cal N}}
\def\cg{{\cal G}}\def\q{\eta}\def\ct{{\cal T}}\def\bbs{\mathbb{S}}
\def\bU{{\mathbf U}}\def\bE{\hat{\mathbf e}}\def\bc{{\mathbf C}}
\def\vs{\varsigma}\def\cg{{\cal G}}\def\ch{{\cal H}}\def\df{\d/\d }
\def\mz{\mathbb{Z}}\def\ms{\mathbb{S}}\def\kb{\hat{\mathbb
K}}\def\cd{\mathcal D}\def\mj{\mathbf{J}}\def\Tr{{\rm Tr}}
\def\bu{{\mathbf u}}\def\by{{\mathrm y}}\def\bp{{\mathbf p}}
\def\k{\kappa} \def\cz{{\mathcal Z}}\def\ma{\mathbf{A}}
\def\me{\mathbf{E}}\def\ra{\mathrm{A}}
\def\ru{\mathrm{u}}\def\rP{\mathrm{P}}\def\rp{\mathrm{p}}\def\z{\zeta}
\def\my{\mathbf Y}\def\ve{\varepsilon}\def\bw{\mathbf
W}\def\hp{\hat{\p}}\def\hh{\hat{h}}\def\hx{\hat{\x}}\def\hk
{\hat{\kappa}}\def\hj{\hat{j}}\def\eb{\mathbf{e}}\def\bj{\mathbf{j}}
\def\tO{\tilde\o}\def\cS{{\cal S}}\def\os{\overline{\s}} \def\K{\hat{K}}
\def\mH{\mathcal{H}} \def\he{\hat{e}}
\def\to{\rightarrow}\def\t{\times}\def\u{\hat e}\def\ep{\epsilon}

\begin{flushright}
 Dedicated to \\{\bf Alexei Sisakian }\\
co-author $\&$ friend
\end{flushright}
\vskip 3cm
\begin{center}
{\huge\bf  On the connection between quantum and classical
descriptions} \vskip 0.5cm {\large\it J.Manjavidze}\vskip 3cm

Joint Institute for Nuclear Research \\(VBLHEP \& LNP, Dubna,
Russia)\\ $\&$\\ Andronikashvili Institute of Physics
\\(LThP, Tbilisi State University, Georgia), \vskip 0.5cm  Tel:
(09621) 6 35 17, Fax: (09621) 6 66 66, E-mail: joseph@jinr.ru
\end{center}
\newpage

\begin{abstract}

{\it The review paper presents generalization of d'Alembert's
variational principle: the dynamics of a quantum system for an
external observer is defined by the exact equilibrium of all acting
in the system forces, including the random quantum force $\hbar j$,
$\forall\hbar$. Spatial attention is dedicated to the systems with
(hidden) symmetries. It is shown how the symmetry reduces the number
of quantum degrees of freedom down to the independent ones. The
sin-Gordon model is considered as an example of such field theory
with symmetry. It is shown why the particles $S$-matrix is trivial in
that model. }

\end{abstract}
\newpage
\tableofcontents \newpage

{\bf Preface}

Present paper is the review of the works which was done after my
first paper \C{yadphys}. I returned from time to time to the idea
\C{yadphys} that it seems interesting to embed the total probability
conservation condition into the quantum field theory formalism and
discuss it with Alexei Sissakian during our team-work. It seems that
this suggestion is unnecessary noting that the $S$-matrix is the
unitary operator and it is not evident why this attempt can give
something new. But it turns out that exist the correspondence among
quantum theory and classics which is independent from the value of
quantum corrections. Besides this new quantum field theory is free
from divergences and the value of quantum corrections ingenuously
depend on the topology of classical field. All that is new from the
point of view of ordinary theory and at last Alexei Sissakian propose
to write on paper all result in details. Present introductory paper
devoted to simplest examples and more interesting field theory models
will be published later.

\newpage

\section{Introduction}\0

The basis of the method of calculations is the following \C{yadphys}.
The $S$-matrix unitarity condition, $S^\dag S=1$, in terms of
amplitudes, $S=1+iA$, looks as follows: \be iA^\dag A=( A-A^\dag).
\l{1.1a}\ee The nonlinearity of this equality points on existence of
the cancelations mechanism (of the real part of amplitude) which
reduces quadratic form down the linear one. Our purpose is to show
how this reduction removes the "unwanted" contributions\footnote{This
means that the theory must be formulated directly in terms of
probability. But notice that it is the frequently used method of
particle physics. For example, one must integrate over unobserved
final state in the inclusive approach to the multiple production
phenomena. Another example: describing the very high multiplicity
(VHM) processes the number of produced particles $n$ must be
considered as the dynamical parameter. In the frame of $S$-matrix
thermodynamics, where the "rough" description of final state is used,
one must also integrate over final particles momenta. In all cases
one must consider quantities $\sim|A|^2$ directly, where $A$ is the
corresponding amplitude.}.

One may consider the simplest vacuum-into-vacuum transition
probability, $|Z|^2$, as the main quantity, where $Z$ is the
functional integral over fields, \be Z=\int D\vp~ e^{iS(\vp)},~
D\vp=\prod_x d\vp(x).\l{Z}\ee One may include into the action,
$S$, also the linear over field $\vp$ term, \be \int dx
J(x)\vp(x)\label{J}\ee to describe production of particles. We
will assume on the early stages that $J=0$. Then the
vacuum-into-vacuum transition probability \be |Z|^2=\int
D\vp_+D\vp^*_-~e^{iS(\vp_+)- iS^*(\vp_-)},\label{Z2}\ee where
$\vp_+$ and $\vp_-$ are completely independent fields.

It can be shown that Eq. (\r{1.1a}) means that a reduced form
must also exist \C{yadphys}: \be |Z|^2= \lim_{j=e=0}
e^{i\kb(j,e)} \int DM e^{iU(\vp,e)},\l{1.5c}\ee where
$\kb=\kb(j,e)$ is a definite differential operator over $j(x)$
and $e(x)$, see the examples (\r{oper}), (\r{2.2z}). The
expansion of $\exp\{i\kb\}$ generates perturbation series. The
functional $U(\vp,e)$ introduces interaction among quantum
degrees of freedom and the integral measure is $\d$-functional:
\begin{equation} DM=\prod_x d\vp(x)\d\le(\f{\d
S(\vp)}{\d\vp(x)}+\hbar j(x)\ri).\label{1.6c}\end{equation}Sometimes the $\d$-like
measure \C{diracmeasure} is called in mathematical literature as the
"Dirac measure". It follows from (\r{1.6c}) that

--- {\it the quantum system for an externa observer looks like
classical which is excited by the external random force $\hbar
j,~\forall\hbar$.}\\ The established generalized correspondence
principle\footnote{This formulation of the principle was
offered by A.Sisakian.} is the main consequence of Eq.
(\r{1.1a}). Therefore the complete set of acceptable field
states for external observer\footnote{Since the "probability"
is considered} is known having (\r{1.6c}).

It is important that the restricted problem is considered. We
will calculate the imaginary part of amplitude believing that
it will be sufficient for us. In this case the unmeasurable
phase of amplitude stay undefined\footnote{ Therewith why must
the calculations of unnecessary, i.e. unmeasurable, phase be
performed? Just in this sense the unitarity condition
(\r{1.1a}) is a necessary one. It says that the real part is
the "unwanted" part of the amplitude.}. A main mathematical
problem in the searching representation (\r{1.5c}) is to find
the way how to find the imaginary part from the modulo square
of amplitude. To be more precise, we will find the imaginary
part as a result of cancelation of "unwanted" contribution in
the modulo square of amplitude.

The $\d$-function (\r{1.6c}) solves the problem of definition
of contributions into the path integral but can not solve the
problem completely since the action of operator $\kb$ remains
unknown. It must be noted that $\exp\{i\kb\}$ generates the
asymptotic series ordinary in quantum theories \C{lipatovv} and
it seems that $\d$-like measure gives nothing
new\footnote{Looking at the approach from the point of view of
the stationary phase methods. In other words, one can think
that the present approach gives nothing new to the Bohr's
correspondence principle.}. But this is not entirely so. I
would like draw attention to the appearance of source of
quantum excitations $\hbar j$ in the r.h.s. of classical
Lagrange equation, i.e. the changes of l.h.s. in equation of
motion leads to the change of $j$. It is crucially important
that (\r{1.6c}) is rightful independently from the value of
$\hbar$.

The theory defined on the Dirac measure (\r{1.6c}) for this
reason has quite unexpected properties, e.g. allows to perform
transformation of the path integral variables. So, it will be
shown that in theories with symmetry the reduced form of
representation (\r{1.5c}) exists: \begin{equation} |Z|^2=
\lim_{j=e=0}e^{i\kb(j,e)}\int DM(j)e^{iU(\vp_c,e)},
\label{1.5c1}\end{equation} where $\kb$ is again the
perturbations generating operator and $U$ introduces
interactions. Note that $\kb$ and $U$ in (\r{1.5c1}) depends on
the sets $\{j_{\x_k},j_{\q_k}\}$, $\{e_{\x_k},e_{\x_k}\}$ of
new variables. One must take this auxiliary variables equal to
zero at the very end of calculations. At the same time the
transformed measure $DM$ is again $\d$-like:
$$ DM=\prod_k\prod_t d\x_k(t)d\q_k(t)\t$$\begin{equation} \t
\d\le(\dot\x_k(t)-\f{\d h}{\d\q_k(t)}-j_{\x_k}(t)\ri) \d\le (\dot
\q_k(t)+\f{\d
h}{\d\x_k(t)}+j_{\q_k}(t)\ri),\label{1.6c1}\end{equation}where $t$ is
the time variable and $h=h(\q)$ is the transformed Hamiltonian:
\begin{equation} h(\q)=H(\vp_c), \label{1.6c2}\end{equation}where
$\vp_c=\vp_c(\bx;\x,\q)$ is given solution of Lagrange equation at
$j=0$.

The formulae (\r{1.6c1}) is the main result. Therefore, as it follows
from it the problem of the quantum field theory with symmetry is
reduced down to quantum mechanical one, with potential defined by
$\vp_c$.

({\bf A}) The Dirac measure (\r{1.6c}) prescribes that $|Z|^2$ is
defined by the $sum$ of $strict$ solutions of equation of motion:
\begin{equation} \f{\d S(\vp)}{\d\vp(x)}=\hbar j(x),\label{eqmot}
\end{equation}in
vicinity of $j=0$, i.e. by definition Eq. (\r{eqmot}) must be solved
expanding the solution over $j$ \footnote{It should be noted that it
may be that the limit $j=0$ is absent. For example it may happen if
the system is unstable against symmetry breaking. This important
possibility will not be considered in present paper.}. Following the
ordinary rule we obviously leave the contribution which ensures the
minimal vacuum energy. On the other hand, having theory on Dirac
measure, which calls for the complete set of contributions, we have
offered another selection rule in our dynamic theory of $S$-matrix.
Namely, we simply propose\footnote{This selection rule is used widely
in classical mechanics, see e.g. formulation of
Kolmogorov-Arnold-Mozer (KAM) theorem \C{KAM}. } that\\
--- {\it the largest terms in the sum over solutions of (\r{eqmot})
are significant from the physics point of view.}\\
To be more precise, this selection rule means that if $G$ is the
symmetry of action and $TG^*$ is the symmetry of the extremum of the
action, then in the situation of a general position only the
trajectories with the highest dimension factor group, $(G/TG^*)$, are
sufficient.

It will be seen that this kind of definition of the "ground state"
extracts the maximally "feeling" symmetry contributions since other
ones will be realized on a zero measure, or, more precisely, only
maximal symmetry breaking field configurations, $\vp_c$, are most
probable. We will call such solution of the problem {\it the field
theory with symmetry}. It is the main formal distinction of present
approach.

It is important here that the zero width of $\d$-function excludes
the interference among contributions from various trajectories.
Therefore the formalism naturally takes into account the
orthogonality of Hilbert spaces built on various trajectories. This
is achieved through the special boundary conditions in the frame of
which the total action of the product $$Z\cdot Z^*=<in|out>\cdot
<out|in>$$ always describes a closed path, i.e. the necessary for
d'Alembert variational principle time reversible motion. It points to
the necessity to be careful with boundary conditions in a considered
formalism\footnote{ The necessity to count all possible boundary
conditions of a given problem was mention to author by L.Lipatov.}.

({\bf B}) The Dowker's theorem \C {dowker} insists that the
semiclassical approximation to be exact for path integrals on the
simple Lie group manifolds. For this reason one can expect that the
quantum-mechanical problems, as well as the field-theoretical ones,
may be at least transparent to the symmetry manifolds.

However we know how to construct correctly the path integral
formalism only in the restricted case of canonical variables \C
{marin}. At first sight the path integrals in terms of generalized
coordinates can be defined through the corresponding transformation.
But there is an opinion that it is impossible to perform the
transformation of path-integral variables: the naive transformation
of coordinates give wrong results because of their stochastic nature
in quantum theories\footnote{One can find corresponding examples in
\C {marin, gul}. The mostly popular method of transformation of the
path-integral variables is a "time-sliced" method \C{hibb}, which
induces corrections to the interaction Lagrangian proportional at
least to $\hbar^2$ \C{groshe}, i.e. the problem of transformation is
of a quantum nature. For this reason the usage of the "time-sliced"
method in general case is cumbersome, see also \C{klein}.}. That is
why such general principle as the conservation of total probability
(\r{1.1a}) should play important role. Indeed, it is evident that
$\d$-like Dirac measure (\r{1.6c}) allows to perform arbitrary
transformation \C {yadphys} just as in the classical mechanics.

Therefore, the theory on Dirac measure straight away leads to the new
for quantum field theory selection rule and latter one gives the
theory with symmetry. All this is attained by transition to the
appropriate variables, $(\x,\eta)\in W$ in our notations. The last
circumstance means that we go away from ordinary spectral analysis of
quantum fluctuations to the description of the classical trajectories
topology conserving deforma\-tions, since $\vp_c=\vp_c(\bx;\x,\q)$ is
given, of symmetry manifold, $W$ \footnote{It will be seen from our
selection rule that the measure on which particles mechanics is
realized is equal to zero in the field theories with symmetry.}. It
must be underlined that our method of transformations is rightful for
arbitrary case, i.e. not only for simple Lie group manifolds, where
the semiclassical approximation is exact.

Next, the dimensions of initial phase space of field and of the
transformed space of independent degrees of freedom, i.e. of the
symmetry manifold, will not coincide. That means that the mapping to
the independent degrees of freedom, $(\x,\eta)$, will be singular.
For this reason the transformation $$\vp_c: \vp\to(\x,\eta)$$ will be
irreversible and the notion of particle should be considered as the
wrong idea of quantum field theory with symmetry
\footnote{Considering gluon production in the frame of Yang-Mills
field theory with symmetry the conclusion that gluons can not be
created should be confirmed by direct calculations, taking into
account also the quark fields. That was mentioned to the author by
P.Culish and will be shown in later publications. It is noticeable
that the mapping in quantum mechanics is not singular and for this
reason both representations before and after transformation have the
equal status.}.

({\bf C}) It will be shown that the result of action of the operator
$\exp\{i\kb\}$ for transformed theories may be expressed as the sum
of contributions on all boundaries $\pa W$: \begin{equation}
|Z|^2=|Z|^2_{sc}+\sum_k\int d\x_k(0)\f{\pa}{\pa\x_k(0)}C_\x+
\sum_k\int d\q_k(0)\f{\pa}{\pa\q_k(0)}C_\q
\label{}\end{equation}where the first term presents a semiclassical
contribution and $C_\x$, $C_\q$ contains quantum corrections. This
result shows that the quantum corrections greatly depend on the
topology of classical trajectory.

This important observation solves a number of problems. For instance,
it is known that the Coulomb trajectory is closed because of
Bargman-Fock symmetry, independent from the initial conditions. For
this reason the corrections on $\pa W$ of Coulomb problem are
canceled and the H-atom problem is pure semiclassical. We will find
the same for sine-Gordon model \C{dash} as the consequence of mapping
on the Arnold's hypertorus \C{arnoldd}.

It is extremely important to keep in mind that the symmetry
constraints can not be taken into account perturbatively over the
interaction constant, $g$. Indeed, we will see below that the
expansion in $polinomial$ theories with symmetry is performed in
terms of the inverse interaction constant, $1/g$. It points to the
absence of the weak-coupling limit in such theories.

In the end our present aim is {\it

--- to find representation (\r{1.5c});

--- to investigate the main properties of theory defined on the Dirac
measure (\r{1.6c});

--- to investigate the structure of perturbation theory generated by
operator $\kb$ on the measure (\r{1.6c1});

--- to find particles production probabilities for theories with
symmetry.}

I understand that the perturbations scheme in terms of new variables,
especially in theories with symmetry, is outside the habitual
one\footnote{See \C{dira, korepinn, jackiw}} and for this reason the
approach will be describe in more details, giving step-by-step the
properties of a new quantization scheme by appropriate examples. I
think that such non-formal scheme of the description is much more
transparent, although the text may contain reiterations with the used
method of description far from completeness.

\newpage

\section{Simplest examples}\0

\subsection{Introduction}

It it has mentioned above a technical aspect of our idea is the
suggestion to calculate the probability without the intermediate step
of calculations of the amplitudes. In present Section we restrict
ourselves to the simplest problem - to the motion of a particle in a
potential $V(x)$.

Let the amplitude $A(x_2,T;x_1,0)$ describes the motion of the
particle from the point $x_1$ to the point $x_2$ during the time $T$.
Using the spectral representation: \begin{equation}
A(x_2,T;x_1,0)=\sum_n
\psi_n(x_2)\psi_n^*(x_1)e^{iE_nT},\label{1.2}\end{equation}we have
for probability: \begin{equation}  W(x_2,T;x_1,0)=\sum_{n_1,n_2}
\psi_{n_1}(x_2) \psi_{n_1}^*(x_1)\psi_{n_2}^*(x_2)\psi_{n_2}(x_1)
e^{i(E_{n_1}-E_{n_2})T}.\label{1.3}\end{equation}Taking into account
the ortho-normalizability condition: \begin{equation}  \int dx
\psi_n(x) \psi_m^*(x)=\d_{n,m},\label{1.4}\end{equation}the total
probability:
\begin{equation}  \int dx_2dx_1 W(x_2,T;x_1,0)=\sum_n\d_{n,n}=\O
\label{inf}\end{equation}is the time independent quantity which
coincides with the number of existing physics states. Therefore, the
amplitude (\r{1.2}) is time dependent, but the total probability
(\r{inf}) is not. This means that the time is the unwanted parameter
from the point of view of experiment described by the probability
(\r{inf}). Notice also the role of boundary condition (\r{1.4}).

The quantity (\r{inf}) is of no interest to experiment. Much more
interesting the probability $\R(E)$, where $E$ is the energy
experimentally measured. The Fourier transform of $A(x_2,T;x_1,0)$
with respect to $T$
\begin{equation}
a(x_2,x_1;E)=\sum_n\f{\psi_n(x_2)\psi_n^*(x_1)}{E-(E_n+i\ve)}\label{1.6}\end{equation}leads
to the probability \begin{equation} \o(x_2,x_1;E)=|a(x_2,x_1;E)|^2=
\sum_{n_1,n_2} \f{\psi_{n_1}(x_2)\psi_{n_1}^*(x_1)}
{E-(E_{n_1}+i\ve)} \f{\psi_{n_2}^*(x_2) \psi_{n_2}(x_1)}
{E-(E_{n_2}-i\ve)}\label{1.7}\end{equation}and the total probability:
$$ \R(E)=\int dx_1dx_2\o(x_2,x_1;E)=\sum_{n}\le| \f{1}{E-E_{n}-i\ve}
\ri|^2=$$\begin{equation} = \f{1}{\ve}\sum_n\Im\f{1} {E-E_{n}-i\ve}=
\f{\pi}{\ve}\sum_n\d(E-E_{n}). \label{1.8}\end{equation}The total
probability $\R(E)$ again coincides with number of existing states
but for all that it is seen that the unphysical, i.e. needless,
states from the point of view of measurement with $E\neq E_n$ was
canseled\footnote{Such contributions enter into the real part of
$a(x_2,x_1;E)$.}.

Let us use now the proper-time representation: \begin{equation} a(x_1
,x_2 ;E)=\sum_{n} \Psi_{n} (x_1)\Psi^{*}_{n} (x_2)i
\int^{\infty}_{0}dTe^{i(E-E_{n}+i\ve)T} \label{6'}
\end{equation}to see the integral form of cancelation of unwanted
contributions and insert it into definition of total probability
($\ve\to+0$):
\begin{equation} \R(E)=\int dx_1 dx_2 |a(x_1 ,x_2 ;E)|^2=\sum_{n}
\int^{\infty}_{0} dT_{+}dT_{-} e^{-(T_{+}+T_{-})\ve}
e^{i(E-E_{n})(T_{+}-T_{-})}. \label{7'} \end{equation}
We  will introduce new time variables instead of $T_{\pm}$:
\begin{equation} T_{\pm}=T\pm\tau, \label{8'} \end{equation}where, as it follows
from Jacobian of transformation, $|\tau|\leq T,~0\leq T\leq \infty$.
But we can put $|\tau|\leq\infty$ since $T\sim1/\ve\rar\infty$ is
essential in integral over $T$. As a result, \begin{equation}
\R(E)=4\pi\sum_{n}\int^{\infty}_{0} dT e^{-2\ve T}
\int^{+\infty}_{-\infty}\f{d\tau}{\pi}
e^{2i(E-E_{n})\tau}=\f{\pi}{\ve}\sum_n\d(E-E_n). \label{9'}
\end{equation}In the last integral all contributions with $E\neq
E_{n}$ has been canceled and only the acceptable from physics point
of view contributions with $E=E_n$ has survived. This peculiarity of
considered interference phenomena which is the consequence of
unitarity condition, i.e. its ability to extract only physics states,
would have the significant applications.

Note also that the product of amplitudes $a\cdot a^*$ was
"linearized" after introduction of "virtual" time $\tau
=(T_{+}-T_{-})/2$, i.e. after transformation (\r{8'}) we start
calculation of the imaginary part. The meaning of such variables will
be discussed also in Sec.2.2.

\subsection{The generalized stationary-phase method}

{\bf 1.} {\it 0-dimensional model}

Let us practise considering the "$0$-dimensional" integral:
\begin{equation}  A= \int^{+\infty}_{-\infty} \frac{dx}{(2\pi)^{1/2}}
e^{i(\frac{1}{2}ax^2 +\frac{1}{3}bx^3)}, \label{11a}
\end{equation}with $\Im a \rightarrow +0$ and $b>0$. This example is
useful since it allows to illustrate practically all technical tricks
of the approach.

We want to compute the "probability" \begin{equation}  R=|A|^2=
\int^{+\infty}_{-\infty}\frac{dx_+ dx_-}{2\pi} e^{i(\frac{1}{2}ax_+^2
+\frac{1}{3}bx_+^3) -i(\frac{1}{2}a^* x_-^2 +\frac{1}{3}bx_-^3)}.
\label{12} \end{equation}New variables: \begin{equation}  x_{\pm}
=x\pm e \label{13}\end{equation}will be introduced to find out the
cancelation phenomenon. In result:
\begin{equation}  R= \int^{+\infty}_{-\infty}\frac{dx de}{\pi} e^{-2(x^2 +e^2 )\Im
a}e^{2i(\Re a\;x +2bx^2)e} e^{2i\frac{b}{3}e^3}, \label{14z}
\end{equation}where the prescription that $\Im a \rightarrow +0$ has been
used. Note that integrations are performed along the real axis.

We  will compute the  integral over $e$ perturbatively. For this
purpose the transformation: \begin{equation}  F(e)=\lim_{j=e'=0}
e^{\frac{1}{2i}\h{j}\hat{e}'} e^{2ije}F(e'), \label{15a}
\end{equation}which is valid for any differentiable function, will be
used. In (\ref{15a}) two auxiliary variables $j$ and $e'$ has been
introduced and the "hat" symbol means the differential over
corresponding quantity:
\begin{equation} \hat{j}=\frac{\partial}{\partial j},\;\;\;
\hat{e'}=\frac{\partial}{\partial e'}. \label{iz} \end{equation}The auxiliary
variables must be taken equal to zero at the very end of
calculations.

Choosing \begin{equation}  \ln F(e)= -2e^2 \Im a +2i\frac{b}{3}e^3
\label{16a} \end{equation}we will find: \begin{equation}
R=\lim_{j=e=0} e^{\frac{1}{2i}\hat{j}\hat{e}}
\int^{+\infty}_{-\infty} dx e^{-2(x^2 +e^2 )\Im
a}e^{2i\frac{b}{3}e^3} \delta (\Re a~x +bx^2+j) . \label{17a}
\end{equation}Therefore, the destructive interference among two exponents in
the product $a\cdot a^*$ unambiguously determines both integrals,
over $x$ and over $e$. The integral over difference $e=(x_+ -x_-)/2$
gives $\delta$-function and then this $\delta$-function defines the
contributions in the  last integral over $x=(x_+ +x_-)/2$. Following
the definition of $\delta$-function only a strict solutions of
equation \begin{equation}  \Re a\;x +bx^2+j=0 \label{18z}
\end{equation}gives the contribution into $R$.

But one can note that this is not the complete solution of the
problem: the expansion of operator exponent $\exp
\{\frac{1}{2i}\hat{j}\hat{e}\}$ generates the asymptotic series. Note
also that it is impossible to remove the source, $j$, dependence
(only harmonic case, $b=0$, is free from $j$).

The equation (\ref{18z}) at $j=0$ has the solutions, at $x_1 =0$ and
at $x_2 =-a/b$. Performing trivial transformation $e\rightarrow ie$,
$\hat{e}\rightarrow -i\hat{e}$ of  auxiliary variable we find at the
limit $\Im a=0$ that the contribution from $x_1$ extremum (minimum)
has the expression\footnote{The contribution of $x_2$ leads to
divergent series.}: \begin{equation} R=\frac{1}{a}
e^{-\frac{1}{2}\hat{j}\hat{e}} (1-4bj/a^2)^{-1/2} e^{2\frac{b}{3}e^3}
\label{19a} \end{equation}and the expansion of an operator exponent
gives the asymptotic  series: \begin{equation}  R=\frac{1}{a}
\sum^{\infty}_{n=0}(-1)^{n} \frac{(6n-1)!!}{n!}
\le(\frac{2b^4}{3a^6}\ri)^n,\;\;\;\;(-1)!!=0!!=1. \label{20a}
\end{equation}This series is convergent in Borel's sense. Therefore
the described destructive interference has not an action upon the
value of perturbation series convergence radii.

Let us calculate now $R$ using stationary phase method. The
contribution from the minimum $x_1$ gives $(\Im a=0)$:
\begin{equation} A=e^{-i\hat{j}\hat{x}} e^{-\frac{i}{2a}j^2}
e^{i\frac{b}{3}x^3} ({i}/{a})^{1/2}. \label{21z} \end{equation}The
corresponding ``probability" is \begin{equation}  R=\frac{1}{a}
e^{-i(\hat{j}_+\hat{x}_+ -\hat{j}_-\hat{x}_-)} e^{-\frac{i}{2a}(j_+^2
-j_-^2)} e^{i\frac{b}{3}(x_+^3 -x_-^3)}. \label{22a}
\end{equation}Introducing new auxiliary variables: \begin{equation}
j_{\pm}=j \pm j_1 ,\;\;\;\;x_{\pm}= x \pm e \label{23z}
\end{equation}and, correspondingly, \begin{equation}
\hat{j}_{\pm}=(\hat{j}\pm\hat{j}_1)/2,\;\;\;\;
\hat{x}_{\pm}=(\hat{x}\pm\hat{e})/2 \label{24z} \end{equation}we find
from (\ref{22a}): \begin{equation}  R=\frac{1}{a}
e^{-\frac{1}{2}\hat{j}\hat{e}} e^{2\frac{b}{3}e^3}
e^{\frac{2b}{a^2}ej^2} \label{25a} \end{equation}This expression does
not coincide with (\ref{19a}) but it leads to the same asymptotic
series (\ref{20a}). We may conclude  that both considered methods of
calculation of $R$ are equivalent since the Borel's regularization
scheme  of asymptotic series gives a unique result.

The difference between this two methods of  calculation is  in
different organization of perturbations. So, if $F(e)$, instead of
(\ref{16a}), is chosen in the form: \begin{equation}  \ln F(e)= -2e^2
\Im a +2i\frac{b}{3}e^3 +2ibx^2 e, \label{26a} \end{equation}we may
find (\ref{25a}) straightforwardly.

Therefore, our method has the freedom in choice of (quantum) source
$j$\footnote{This freedom was mentioned firstly by A.Ushveridze.}.
Indeed, the transition from perturbation theory with Eq.(\ref{16a})
to the theory with Eq.(\ref{26a}) formally looks like following
transformation of the argument of $\d$-function: \begin{equation}
\delta (ax+bx^2 +j)= \lim_{e'=j'=0}e^{-i\hat{j}'\hat{e}'} e^{i(bx^2
+j)e'} \delta (ax +j'). \label{27a} \end{equation}Here the
transformation (\ref{15a}) of the Fourier image of $\d$-function was
used. Inserting Eq.(\ref{27a}) into (\ref{17a}) we easily find
(\ref{25a}).

During analytic calculations it will be useful to have a
corresponding quantum sources of the new dynamical variables.
Formally this will be done using transformation (\r{27a}). Note that
this transformation will not lead to changing of the Borel's
regularization procedure.

{2.} {\it 1-dimensional model}

Let us calculate now the probability using the path-integral
definition of amplitudes \C{yadphys}. Calculating the quantity
\begin{equation} |A|^2=\rm<in|out><in|out>^*=<in|out><out|in>,
\label{2.1z}\end{equation}the converging and diverging waves in the
product $A\cdot A^*$ interfere in such a way that the continuum of
contributions cancel each other. Indeed, the amplitude
\begin{equation} A(x_2,T;x_1,0)=\int^{x(T)=x_2}_{x(0)=x_1}
\f{Dx}{C_T}
e^{-iS_T(x)},~Dx=\prod_{t=0}^T\f{dx(t)}{(2\pi)^{1/2}},\label{integ}\end{equation}where
the action $S_T$ is given by the expression: \begin{equation}
S_T(x)=\int^T_0dt \le(\f{1}{2}~\dot
x^2-v(x)\ri)\label{2.3z}\end{equation}and $C_T$ is the standard
normalization coefficient: \begin{equation} C_T=\int^{x(T)=
x_2}_{x(0)=x_1}Dx e^{\f{i}{2}\int^T_0dt~\dot
x^2}\label{2.4}\end{equation}Let us calculate the quantity
\begin{equation} R(x_2,T;x_1,0)=\int^{x_\pm(T)=x_2}_{x_\pm(0)=x_1}
\f{Dx_+}{C_T}\f{Dx_-}{C_T^*}
e^{-iS_T(x_+)+iS_T(x_-)}\label{2.5b}\end{equation}We assume for
simplicity that the integration in (\r{integ}) is performed over real
trajectories. Later a general case of complex trajectories will be
considered.

The convergence of functional integral at that is not important.
One may restrict the range of integration for better confidence,
or introduce into the Lagrangian $i\ve$ term, and later remove the
restriction in the expression (\r{2.11c}). It is interesting that
the interference phenomena naturally regularize divergent
integrals of (\r{integ}) type, accumulating divergence into
$\d$-function.

In order to take into account explicitly the interference between
contributions of the trajectories $x_+(t)$ and $x_-(t)$ we shall go
over from the integration over two independent trajectories $x_+$ and
$x_-$ to the pair $(x,e)$: \begin{equation}  x_\pm(t)=x(t)\pm
e(t).\label{2.6}\end{equation}It must be stressed that the
transformation (\r{2.6}) is linear and for this reason may be done in
the path integral. Substituting (\r{2.6}) into (\r{2.5b}) the
argument of the exponent takes the form \begin{equation}
S_T(x+e)-S_T(x-e)=2\int_0^Tdt e(\ddot x+v'(x))-
U_T(x,e),\label{2.7b}\end{equation}where $U_T(x,e)$ is the remainder
of the expansion in powers of $e(t)$ ($U_T=O(e^3)$). Note that in
(\r{2.7b}) we have discarded the "surface" term
\begin{equation} \int_0^Tdt\pa_t(e\dot x)=e(T)\dot x(T)-e(0)\dot
x(0)=0,\label{2.8z}\end{equation}since the boundary points of the
trajectories $x_+(0)=x_-(0)=x_1$ and $x_+(T)=x_-(T)=x_2$ are not
varied, i.e. \begin{equation}
e(0)=e(T)=0.\label{2.9b}\end{equation}Next,
\begin{equation}  Dx_+Dx_-=J Dx De=2\pi J\prod_{t=0}^T
dx(t)\prod_{t\neq0,T}\f{de(t)}{2\pi},\label{2.10}\end{equation}where $J$ is an
unimportant Jacobian of the transformation.

As a result of the replacement (\r{2.6}) we have \begin{equation}
R(x_2,T;x_1,0)=2\pi J \int^{x(T)=x_2}_{x(0)=x_1}
\f{Dx}{|C_T|^2}\int^{e(T)=0}_{e(0)=0} De~e^{2i\int_0^T dt e(\ddot
x+v'(x))+U_T(x,e)}. \label{2.11c}\end{equation}One can make use of
the formulae
\begin{equation} e^{iU_T(x,e)}=
e^{\kb(e',j)}e^{iU_T(x,e')}e^{-2i\int_0^Te(t)
j(t)dt},\label{2.12}\end{equation}where we have introduced the operator
\begin{equation}
\kb(e,j)=\lim_{e=j=0}\exp\le\{-\f{1}{2i}\int_0^T\f{\d}{\d j(t)}
\f{\d}{\d e(t)}\ri\},\label{oper}\end{equation}after which from
(\r{2.11c}) we have found that
$$ R(x_2,T;x_1,0)=2\pi Je^{\kb(e',j)}\int^{x(T)=x_2}_{x(0)=x_1}
\f{Dx}{|C_T|^2} e^{iU_T(x,e')}\t$$
$$\t\int^{e(T)=0}_{e(0)=0} De~\exp\le\{2i\int_0^T dt (\ddot
x+v'(x)-j)e\ri\}=$$\begin{equation} =2\pi J e^{\kb(e,j)}\int^{x(T)=
x_2}_{x(0)=x_1} \f{Dx}{|C_T|^2} e^{iU_T(x,e)}\prod_{t\neq
0,T}\d(\ddot x+v'(x)-j),\label{resul}\end{equation}where the
functional $\d$-function \begin{equation}  \prod_{t\neq 0,T}\d(\ddot
x+v'(x)-j)= \int^{e(T)=0}_{e(0)=0} De~\exp\le\{2i\int_0^T dt (\ddot
x+v'(x)-j)e\ri\}\label{2.15}\end{equation}has arisen as a result of
total reduction of unnecessary contributions from the point of view
of equation of motion \begin{equation}  \ddot
x(t)+V'(x)=j(t).\label{2.16b}\end{equation}The operator (\r{oper}) is
Gaussian so that the system is perturbed by the random force $j(t)$.

If $x(t)$ is the "true" trajectory and the virtual deviation is
$e(t)$ then the quantity $e(\ddot x+v'(x)-j)$ coincides with the
virtual work. It must be equal to zero in classical mechanics
since only the time reversible motion is considered. In result we
came to equation of motion since $e$ is arbitrary in classics.

The difference $S_T(x_+)-S_T(x_-)$ in (\r{2.5b}) with boundary
conditions (\r{2.9b}) coincides with the action of reversible motion.
Upon the substitution (\r{2.6}) we have identified the mean
trajectory, $x(t)$, and the deviation from it, $e(t)$. One must
integrate over $e(t)$ in quantum case, in contrast to classical one.
In result the measure of the remaining path integral over mean
trajectory $x(t)$ takes the Dirac $\d$-function form which
unambiguously chooses the "true" trajectory.

In other words, the proposed definition of the measure of the path
integral is generalization of classical d'Alambert's principle on the
quantum case. The theory in the frame of this principle can take into
account any external perturbations, $j(t)$ in our case, if the time
reversibility of motion is conserved. In quantum case the
reversibility is established through the boundary conditions
(\r{2.9b}). Next, one may generalize the approach adding also the
probe force which can lead to dynamical symmetry breaking
\C{weinb}\footnote{It is important that if the expectation value of
the probe force is not equal to zero then the symmetry is broken.
This important possibility will not be considered in present work.}.

In the semiclassical approximation $\kb(e,j)=1$ and taking the limit
$e=j=0$ we find that \begin{equation}  R(x_2,T;x_1,0)=2\pi
J\int^{x(T)=x_2}_{x(0)=x_1} \f{Dx}{|C_T|^2}\prod_{t\neq 0,T}\d(\ddot
x+v'(x)),\label{2.17}\end{equation}Let the solution of the
homogeneous equation
\begin{equation} \ddot x+v'(x)=0\label{2.18b}\end{equation}be $x_c(t)$, with
$x_c(0)=x_1$ and $x_c(T)=x_2$. Then \begin{equation}
R(x_2,T;x_1,0)=2\pi J\int^{x(T)=x_2}_{x(0)=x_1}
\f{Dx}{|C_T|^2}\prod_{t\neq 0,T}\d(\ddot
x+v''(x_c)x),\label{2.19b}\end{equation}The remaining integral is
calculated by the standard methods\footnote{Here it is more
convenient to represent (\r{2.19b}) as a production of two Gauss
integrals; later on more effective method of calculation of the
functional determinant will be offered.}. As a result we find
\begin{equation}  R(x_2,T;x_1,0)= \f{1}{2\pi}\le|\f{\pa^2
S_T(x_c)}{\pa x_c(0)\pa x_c(T)}\ri|_{x_c(0)=
x_1,x_c(T)=x_2}.\label{2.20}\end{equation}Next, let us recall that
the full derivative of the classical action is \begin{equation}
dS=p_2dx_2-p_1dx_1,\label{2.21}\end{equation}where $p_2$ and $p_1$
are, respectively, the final and initial momentum. Noting this
definition,
\begin{equation}  \le|\f{\pa^2 S_T}{\pa x_1\pa x_2}\ri|dx_2=dp_1,\label{2.22z}\end{equation}and
in result we find that \begin{equation} \int dx_1 dx_2
R(x_2,T;x_1,0)=\int\f{dx_1
dp_1}{2\pi}=\O^2,\label{2.23b}\end{equation}which coincides with
(\r{inf}), i.e. it agree with conservation of total probability since
(\r{2.23b}) again coincides with the total number of physical states.

Deriving (\r{2.23b}) we somewhat simplify the problem considering a
unique solution of Eq.(\r{2.18b}). A more complicate and important
examples will be considered in the next Sections.

\subsection{Complex trajectories}

Let us consider the one dimensional motion with fixed energy $E$ on
the complex trajectory\footnote{The necessity to extend the formalism
on the case of complex trajectories was mention to the author by
A.Slavnov.}. The corresponding amplitude has the form:
\begin{equation}  A(x_1 ,x_2 ;E)= i\int^{\infty}_{0} dT e^{iET}
\int_{x_1 =x(0)}^{x_2=x(T)} D_{C_+}x e^{iS_{C_+}(x)}, \label{28a}
\end{equation}where the action \begin{equation}
S_{C_+}(x)=\int_{C_+} dt (\frac{1}{2}\dot{x}^2 -v(x)) \label{29z}
\end{equation}and the measure \begin{equation}  D_{C_+}x=\prod_{t\in
C_+}\frac{dx(t)}{(2\pi)^{1/2}} \label{30z} \end{equation}are defined
on the shifted in the upper half time plane Mills' contour $C_+
=C_+(T)$ \C{millss}:
\begin{equation}  t\rightarrow t+i\varepsilon, \;\;\;\varepsilon \to+0, \;\;\;0\leq
t \leq T. \label{iiz} \end{equation}Therefore, we will consider
integration over real functions of complex variables:
\begin{equation}  x^*(t)=x(t^*).\label{condi}\end{equation}It must be
underlined also that the boundary conditions in (\r{28a}) have the
classical meaning, i.e. they do not vary, and $x_1$, $x_2$ are the
real quantities.

The probability looks as follows: $$ R(E)= \int^{\infty}_{0}
e^{iE(T_+ -T_-)} \int^{x_\pm(T_\pm)=x_2}_{x_\pm(0)=x_1} D_{C_+}x_+
D_{C_-}x_-\t
$$\begin{equation} \t e^{iS_{C_+ (T_+ )}(x_+ ) - iS_{C_- (T_- )}(x_-
)}, \label{31a} \end{equation}where $C_- (T)=C^{*}_{+}(T)$ is the time contour
in the lower half of complex time plane.

New time variables \begin{equation}  T_{\pm}=T\pm\tau \label{32a}
\end{equation}will be used. Considering $\Im E\rightarrow +0$ we can
consider $T$ and $\tau$ as the independent variables:
\begin{equation}  0\leq T \leq \infty,\;\;\; -\infty \leq \tau \leq
\infty. \label{33z} \end{equation}We will apply the boundary
conditions, see (\r{31a}): \begin{equation} x_1=x_+(0) =x_-(0),~~
x_2=x_+(T_+) =x_-(T_-).\label{34a} \end{equation} Inserting (\r{32a})
one can find in zero order over $\tau$ from (\r{34a}) that
\begin{equation}  x_+(0) =x_-(0),~~ x_+(T) =x_-(T),
\label{34b}\end{equation}Now we will introduce also the mean trajectory
$x(t)=(x_+ (t) +x_- (t))/2$ and the deviation $e(t)$ from $x(t)$:
\begin{equation} x_{\pm}(t)=x(t)\pm e(t). \label{35a}\end{equation}We have consider
$e(t)$ and $\tau$ as the virtual quantities. The integrals over $e$
and $\tau$ will be calculated perturbatively. In zero order over $e$
and $\tau$, i.e. in the semiclassical approximation, $x$ is the
classical path and $T$ is the total time of classical motion. Note
that one can do surely the linear transformations (\ref{35a}) in the
path integrals.

The higher terms over $\tau$ put a unphysical constrains on the
trajectory $x(t)$: $$\f{d^{(2n+1)}x(T)}{dT^{(2n+1)}}=0,~
n=0,1,2,...,$$ since $e(t)$ must be arbitrary. Therefore, to avoid
this constraints and since the boundaries have classical unvaried
meaning we will use the minimal boundary conditions: \begin{equation}
e(0)=e(T)=0, \label{36a}\end{equation}which ensures the time
reversibility. Note that it is sufficient to have (\r{36a}) if the
integrals over $e(t)$ are calculated perturbatively. At the same time
\begin{equation} x(0)=x_1,~x(T)=x_2.\label{36x}\end{equation}
Let us extract now the linear over $e$ and $\tau$ terms from the
closed-path action: $$ S_{C_+ (T_+ )}(x_+ ) - S_{C_- (T_- )}(x_- )=
$$\begin{equation} =-2\tau H_{T}(x)- \int_{C^{(+)}(T)}dt e(\ddot{x}
+v'(x))- \tilde{H}_T (x;\tau)-U_T (x,e), \label{37a} \end{equation}where \begin{equation}
C^{(+)}(T)=C_+ (T)+C_-(T) \label{iiiz} \end{equation}is the total-time path, $H_T$
is the Hamiltonian: \begin{equation}  2H_T (x)=-\frac{\partial}{\partial T} (S_{C_+
(T)}(x) + S_{C_- (T)}(x)), \label{38z} \end{equation}and \begin{equation}  -\tilde {H}_T (x;\tau
)= S_{C_+ (T+\tau)}(x) - S_{C_- (T-\tau )}(x)+ 2\tau H_T (x), \label{39z}
\end{equation}
\begin{equation}  -U_T (x,e)=S_{C_+ (T)}(x+e)-S_{C_- (T)}(x-e)+ \int_{C^{(+)}}
dt e(\ddot{x}+v'(x)) \label{40z} \end{equation}are the remainder terms, where
$v'(x)=\partial v(x)/\partial x$.  Deriving the decomposition
(\ref{37a}) the definition \begin{equation}  C_- (T)=C^* _+ (T) \label{41z} \end{equation}and
the boundary conditions (\ref{36a}) was used.

One can find the compact form of expansion of $$ e^{-i\tilde{H}_T
(x;\tau)-iU_T (x,e)}$$ over $\tau$ and $e$ using formulae ({\r{15a}):
$$ \exp\{-i\tilde{H}_T (x;\tau)-iU_T (x,e)\}=
\exp\le\{\frac{1}{2i}\hat{\omega}\hat{\tau}'- i\int_{C^{(+)} (T)}dt
\hat{j}(t)\hat{e}'(t)\ri\}\t $$\begin{equation}  \t \exp\le\{
2i\omega\tau+i\int_{C^{(+)} (T)}dt j(t)e(t) \ri\} \exp\{-i\tilde{H}_T
(x;\tau')-iU_T (x,e')\}. \label{42a} \end{equation}At the end of
calculations the auxiliary variables $(\omega,\tau',j,e')$ should be
taken equal to zero.

Using (\ref{37a}) and (\ref{42a}) we find from (\ref{31a}) that $$
R(E)=2\pi \int^{\infty}_{0}dT
\exp\le\{\frac{1}{2i}\hat{\omega}\hat{\tau}- i\int_{C^{(+)} (T)}dt
\hat{j}(t)\hat{e}(t)\ri\}\times $$\begin{equation}  \times \int Dx
\exp\{-i\tilde{H}_T (x;\tau)-iU_T (x,e)\} \delta (E+\omega -H_T
(x))\prod_{C^{(+)}} \delta (\ddot{x}+v'(x)-j). \label{43a}
\end{equation}The expansion over the differential operators:
\begin{equation} \frac{1}{2i}\hat{\omega}\hat{\tau}-i\int_{C^{(+)}
(T)}dt \hat{j}(t)\hat{e}(t)
=\frac{1}{2i}\le(\frac{\partial}{\partial\omega}\frac{\partial}{\partial
\tau} +\Re\int_{C+}dt\frac{\delta}{\delta j(t)}\frac{\delta}{\delta
e(t)}\ri) \label{44a} \end{equation}will generate the perturbation series. We
propose that it is summable in Borel sense.

The first $\delta$-function in (\ref{43}) fixes the conservation of
energy: \begin{equation}  E+\omega =H_T (x) \label{45za}
\end{equation}where $E$ is the observed energy, $H_T (x)$ is the
energy at the mean trajectory at the time moment $T$ and $\omega$ is
the energy of quantum fluctuations. The second
$\delta$-function\footnote{Following shorthand entry of $\d$-function
of the complex argument: $\prod_{C^{(+)}}
\d(f(t))=\prod_{C_+}\d(f(t)) \prod_{C_-}\d(f(t))=\prod_{C_+}\d(\Re
f(t)+i\Im f(t))\d(\Re f(t)-i\Im f(t))=\prod_{C_+}\d(\Re
f(t))\cdot\\\d(\Im f(t))$ will be useful during calculations. The
condition (\r{condi}) is important here. The inessential constant can
be canceled by normalization. So, in the result of analytical
continuation of $C_\pm$ on the real axis the product of two
$\d$-functions reduces to single one since $\d^2(\Re
f(x))=\d(0)\d(\Re f(x))=\d(0)\d(f(x))$ and $\d(0)$ must be canceled
by normalization. Offered abbreviated notation will allow to consider
$\d$-function on the complex time contour as the ordinary one.}
$$ \prod_{t\in C^{(+)}} \d (\ddot{x}+v'(x)-j)= (2\pi )^2 \int
\prod_{t\in C^{(+)}}\f{de(t)}{\pi}
\d(e(0))\d(e(T))\t$$\begin{equation}  \t e^{-2i\Re\int_{C_+}dt
e(\ddot{x}+v'(x)-j)}= \prod_{t\in C_+ (T)} \d (\Re
(\ddot{x}+v'(x)-j))\d(\Im (\ddot{x}+v'(x)-j)) \label{46a}
\end{equation}fixes the function $x(t)$ of complex argument on
$C^{(+)}$ completely by the equation \begin{equation}
\ddot{x}+v'(x)=j.\label{48a}\end{equation}The physics meaning of
$\d$-function (\ref{46a}) was discussed in Sec.2.3 noting that the
unitarity condition of quantum theories plays the same role as the
d'Alambert's variational principle in classical mechanics.

In (\r{48a}) $j(t)$ describes the external quantum force. The
solution $x_j (t)$ of  this equation will be found  expanding it over
$j(t)$: \begin{equation}  x_j (t)=x_c (t)+\int dt_1 G(t,t_1 )j(t_1
)+... \label{49a} \end{equation}This is sufficient since $j(t)$ is
the auxiliary variable\footnote{See also footnote 15.}. In this
decomposition $x_c (t)$ is the strict solution of unperturbed
equation: \begin{equation} \ddot{x}+v'(x)=0 \label{50a}
\end{equation}Note that the functional $\delta$-function in
(\ref{46a}) does not contain the end-point values of $x(t)$, at $t=0$
and $t=T$. This means that if we integrate over $x_1$ and $x_2$ then
the initial conditions to the Eq.(\ref{50a}) are not fixed and the
integration over them must be performed.

Inserting (\ref{49a}) into (\ref{48a}) we find the equation for Green
function: \begin{equation}  (\partial^2 +v''(x_c))_t
G(t,t';x_c)=\delta (t-t'). \label{51b} \end{equation}It is too hard
to find the exact solution of this equation if $x_c (t)$ is the
nontrivial function of $t$. We will see that the canonical
transformation to the (action-angle)-type variables can help to avoid
this problem, see following Section.

\subsection{Conclusions}

1. The path integral must be defined on the Mills time contour. This
condition will be important in the field theories with high
space-time symmetries (such as the Yang-Mills type theory) since it
seems that for such theories with symmetry one can not perform surely
the analytic continuation over time variable\footnote{The fact that a
theory must satisfy certain conditions upon analytic continuation
over time variable is clear from \C{af}.}.

2. The quantization can be performed without transition to the
canonical formalism, using only the Lagrange one which is more
natural for relativistic field theories.

3. Only the exact solutions of the equation of motion must be taken
into account defining the contributions into the functional integral.

\newpage

\section{Path integrals on Dirac measure}\0

\subsection{Introduction}

In present Section we will offer two methods which may simplify
calculation of path integrals on Dirac measure. They are based on the
possibility to perform transformation of the path-integral variables.

We will consider two examples. In the first example the
transformation to the (action,angle)-type variables will be
considered. This example shows how much the calculations of path
integrals may be simplified.

In the second part of present Section the coordinate transformation
will be described. For the sake of definiteness the transformation to
cylindrical coordinates will be considered.

\subsection{Canonical transformation}

Let us introduce the first-order formalism. We will insert in
(\ref{43a}) \begin{equation}  1=\int Dp \prod_{t} \delta
(p-\dot{x}).\label{52z}\end{equation}As a result, $$ R(E)=2\pi
\int^{\infty}_{0}dT e^{\f{1}{2i}(\hat{\o}\hat{\tau}+\Re\int_{C_+
(T)}dt \hat{j}(t)\hat{e}(t))}\int Dx Dp e^{-i\tilde{H}_T
(x;\tau)-iU_T (x,e)}\t $$\begin{equation} \t \d (E+\omega -H_T (x))
\prod_{t}\d\le(\dot{x}-\frac{\partial H_j}{\partial p}\ri)
\delta\le(\dot{p}+\frac{\partial H_j}{\partial x}\ri), \label{53a}
\end{equation}where \begin{equation}  H_{j}=\frac{1}{2}p^2 +v(x)-jx
\label{54a} \end{equation}may be considered as the total Hamiltonian
which is time dependent through $j(t)$. Notice that in present
simplest case $x$ and $p$ are independent parameters and therefore
(\r{54a}) define the Hamiltonian.

Instead of pare $(x(t),p(t))$ we introduce new pare $(\th(t),h(t))$
inserting in (\ref{53a}) \begin{equation}  1=\int \prod_{t}d\theta
dh\delta\le(h-\frac{1}{2}p^2 -v(x)\ri)\delta\le(\theta -\int^{x}dx
(2(h-v(x)))^{-1/2}\ri). \label{55a} \end{equation}Note that the
integral measures in (\ref{53a}) and (\ref{55a}) are both $\d$-like,
i.e. have the equal power. It allows to change the order of
integration and firstly integrate over $(x,p)$. We find that $$
R(E)=2\pi \int^{\infty}_{0}dT
e^{\f{1}{2i}(\hat{\o}\hat{\tau}+\Re\int_{C_+ (T)}dt
\hat{j}(t)\hat{e}(t))}\int D\th Dh e^{-i\tilde{H}_T (x_c;\tau)-iU_T
(x_c,e)}\t $$\begin{equation}  \t \d (E+\omega -h(T))
\prod_{t}\d\le(\dot{\theta}-\frac{\partial H_c}{\partial h}\ri)
\d\le(\dot{h}+\frac{\partial H_{c}}{\partial \theta}\ri), \label{56a}
\end{equation}where \begin{equation}  H_c =h-jx_c (h,\theta) \label{57}
\end{equation}is the transformed Hamiltonian and $x_c (\theta,h)$ is the given
solution of algebraic equation: \begin{equation}  \theta= \int^{x}dx
(2(h-v(x)))^{-1/2}, \label{}\end{equation}i.e. $x_c$ is the classical
trajectory parametrized in terms of $h(t)$ and $\theta(t)$.

As it follows from (\r{56a}) new variables, $h(t)$ and $\th(t)$, are
subjected to the action of quantum force $j(t)$ and the topology of
classical trajectory $x_c$ remains unchanged.

So, instead of Eq.(\ref{48a}) we must solve the equations:
\begin{equation} \dot{h}=j\frac{\partial x_c}{\partial \theta},
\;\;\;\;\;\dot{\theta}=1-j\frac{\partial x_c}{\partial h}, \label{58a}
\end{equation}which have a simpler structure. Expanding the solutions over $j$
we will find the infinite set of recursive equations. This is the
important peculiarity of used quantization scheme.

Note now that $j\partial x_c/\partial \theta$ and $j\partial
x_c/\partial h$ in the r.h.s. can be considered as the new sources.
We will use this property of Eqs.(\ref{58a}) and introduce in the
perturbation theory new "renormalized" sources: \begin{equation}  j_h
=j\frac{\partial x_c}{\partial \theta},
\;\;\;\;\;j_{\theta}=j\frac{\partial x_c}{\partial h},
\label{59z}\end{equation}i.e. $j_\x$ and $j_\q$ are the forces on the
cotangent bundle. We will use transformation (\ref{27a}):
\begin{equation}  \prod_{t}\delta (\dot{h}-j\frac{\partial
x_c}{\partial \theta})= e^{\f{1}{2i}\Re\int_{C_+}dt\hat{j}_h
(t)\hat{e}_h (t)} e^{2i\Re\int_{C_+}e_h j\f{\pa x_c}{\pa \th}}
\prod_{t}\d (\dot{h}-j_h) \label{60a} \end{equation}and
\begin{equation} \prod_{t}\delta (\dot{\theta}-1+j\frac{\partial
x_c}{\partial h})= e^{\f{1}{2i}\Re\int_{C_+}dt\hat{j}_{\theta}
(t)\hat{e}_{\theta} (t)} e^{2i\Re\int_{C_+}e_{\theta} j\frac{\partial
x_c}{\partial h}} \prod_{t}\d (\dot{\theta}-1-j_{\theta})
\label{61b}\end{equation}to introduce them. The re-scaling of source
$j$ lead to the re-scaling of auxiliary field $e$. In the new
perturbation theory we will have two sources $j_h$, $j_{\theta}$ and
two auxiliary fields $e_h$, $e_{\theta}$. Notice that the momentum
$p$ never arose.

Inserting (\ref{60a}), (\ref{61b}) into (\ref{56a}) we find: $$
R(E)=2\pi \int^{\infty}_{0}dT e^{\frac{1}{2i}(\hat{\omega}\hat{\tau}-
i\int_{C^{(+)}}dt(\hat{j}_h (t)\hat{e}_h (t)+ \hat{j}_{\theta}
(t)\hat{e}_{\theta} (t)))}\t$$$$\t\int Dh D\theta e^{ -i\tilde{H}_T
(x_c ;\tau)-iU_T (x_c ,e_c )}\t$$ \begin{equation} \t \delta
(E+\omega -h(T)) \prod_{t}\delta (\dot{\theta}-1-j_{\theta}) \delta
(\dot{h}-j_h ), \label{62} \end{equation}where \begin{equation}  e_c
= e_h \frac{\partial x_c}{\partial \theta} -e_{\theta} \frac{\partial
x_c}{\partial h} \label{63}\end{equation}carry the simplectic
structure of Hamilton equations of motion and the "hat" symbol means
differential operator over corresponding quantity. At the very end
one should take all auxiliary variables, $(e_h,j_h,e_\th,j_\th)$,
equal to zero.

Hiding the $x_c (t)$ dependence into $e_c$ we solve the  problem of
the functional determinants, see (\ref{62}), and simplify the
Hamilton equations of motion as much as  possible: \begin{equation}
\dot{h}(t)=j_h (t),\;\;\;\;\;\dot{\theta}(t)=1+j_{\theta}(t)
\label{64}
\end{equation}We will use the boundary conditions \begin{equation}  h(0)=h_0
,\;\;\;\;\theta (0)=\theta_0, \label{65} \end{equation}as the
extension of boundary conditions in (\ref{31a}). This lead to the
following Green function of transformed perturbation theory:
\begin{equation} g(t-t')=\Theta (t-t'), \label{66} \end{equation}with the
properties of projection operator: \ba \int dt dt' g^2 (t-t')=\int dt
dt' g(t-t'),
\n \\
\int dt dt' g(t-t') g(t'-t)=0 \label{66a} \ea and, at the same time,
we will assume that \begin{equation}   g(0)=1. \label{66b}
\end{equation}It is important to note that $\Im g(t)$ is regular on
the real time axis. This  is the very simplification of the
perturbation theory since it eliminates the doubling of degrees of
freedom. One may use here the analytical continuation to the real
time axis.

In result, shifting $C_+$ and $C_-$ contours on the real time axis we
find: $$ R(E)=2\pi \int^{\infty}_{0}dT
e^{\frac{1}{2i}(\hat{\omega}\hat{\tau}+ \int^\infty_0 dt_1 dt_2
\Theta (t_1 -t_2 ) (\hat{e}_h (t_1 )\hat{h}(t_2 )+
\hat{e}_{\theta}(t_1 )\hat{\theta} (t_2 )))}\t $$\begin{equation}  \t
\int dh_0 d\theta_0 e^{-i\tilde{H}_T (x_c ;\tau)-iU_T (x_c ,e_c )}
\delta (E+\omega -h_0 +h(T)), \label{67} \end{equation}where the
solutions of eqs.(\ref{64})  was used. In this expression $x_c
(t)=x_c (h_0 -h(t), t+\theta_0 -\theta (t))$ and $(h(t),e_h (t),
\theta (t),e_{\theta}(t))$ are the auxiliary fields. At the very end
one must take them equal to zero.

\subsection{Selection rule}

Let us consider the theory with Lagrangian \begin{equation}
L(x)=\f{1}{2} \dot{x}^2
-\f{1}{2}\o^2x^2-\f{g}{4}x^4.\label{2.81}\end{equation}The Dirac
measure gives the equation (of motion): \begin{equation}
\ddot{x}+\o^2x+gx^3=j. \label{2.82}\end{equation}It has two
solutions:
\begin{equation} x_1(t)=x_c(t)+O(j),~~ x_2(t)=O(j).\label{2.83}\end{equation}For
this reason \begin{equation}  R(E)=R_1(E;x_1)+
R_2(E;x_2)\label{2.84}\end{equation}and which one defines $R(E)$ is a
question. Following to our selection rule just $R_1$.  This will be
shown.

Let us return now to the example with Lagrangian (\r{2.81}). In the
semiclassical approximation \begin{equation} R_1(E;x_1)=\int_0^\infty
dT\int_0^\infty dh_0\int_{-\infty}^{+\infty} d\theta_0
e^{-iU_T(x_c,0)}\d(E-h_0).\label{}\end{equation}Therefore,
\begin{equation} R_1(E;x_1)\sim \int_{-\infty}^{+\infty}
d\theta_0\equiv\O,\label{}\end{equation}i.e. it is proportional to the volume of
group of time translations.

At the same time \begin{equation}
R_2(E;x_2)=O(1)\label{}\end{equation}in the semiclassical
approximation. Therefore, \begin{equation}
R=R_1(1+O(1/\O)).\label{}\end{equation}This result explains the
source of chosen selection rule.

\subsection{Coordinate  transformation}

In this section the coordinate transformation of two dimensional
quantum mechanical model with potential \begin{equation}
v=v((x^{2}_{1}+x^{2}_{2})^{1/2}) \label{69} \end{equation}will be
considered. Repeating calculations of previous sections,
\begin{equation} R(E)=2\pi \int^{\infty}_{0}dT
e^{\frac{1}{2i}\hat{\omega}\hat{\tau}- i\int_{C^{(+)} (T)}dt
\hat{\vec{j}}(t)\hat{\vec{e}}(t)} \int D^{(2)}M(x) e^{-i\tilde{H}_T
(x;\tau)-iU_T (x,e)}, \label{70} \end{equation}where the $\d$-like Dirac measure
\begin{equation}  D^{(2)}M(x) = \delta (E+\omega -H_T (x))\prod_t d^2
x(t)\d^{(2)}(\ddot{x}+v'(x)-j). \label{71} \end{equation} In the
classical mechanics the problem with potential (\ref{69}) is solved
in the cylindrical coordinates: \begin{equation}  x_1 =r
\cos\phi,\;\;\;\;\;x_2 =r \sin\phi. \label{72} \end{equation}We
insert into (\ref{70}) \begin{equation}  1=\int Dr D\phi \prod_{t}
\delta (r-(x^{2}_{1}+x^{2}_{2})^{1/2}) \delta (\phi -\arctan
\frac{x_2}{x_1}). \label{73} \end{equation}to perform the
transformation. Note that the transformation (\ref{72}) is not
canonical. In result we will find a new measure: \begin{equation}
D^{(2)}M(r, \phi ) =\delta (E+\omega -H_T (x))\prod_t dr d\phi
J(r,\phi ), \label{74}
\end{equation}where the Jacobian of transformation \begin{equation}
J(r,\phi )=\int \prod d^2 x \delta^{(2)} (\ddot{x}+v'(x)-j) \delta
(\phi -\arctan \frac{x_2}{x_1}) \delta
(r-(x^{2}_{1}+x^{2}_{2})^{1/2}) \label{75} \end{equation}is the
product of two $\delta$-functions: \begin{equation} J(r,\phi)=\prod_t
r^2 (t) \delta (\ddot{r}-\dot{\phi}^2 r+v'(r)-j_r) \delta(\partial_t
(\dot{\phi}r^2 )-rj_{\phi}), \label{76} \end{equation}where
$v'(r)=\partial v(r)/\partial r$ and
\begin{equation}  j_r =j_1 \cos\phi +j_2
\sin\phi,\;\;\;\;j_{\phi}=-j_1 \sin\phi +j_2\cos\phi \label{77} \end{equation}are
the components of $\vec{j}$ in the cylindrical coordinates.

It is useful to organize the perturbation theory in terms of $j_r$
and $j_{\phi}$. For this purpose following transformation of
arguments of $\delta$-functions  will be used: \begin{equation}
\prod_t \delta (\ddot{r}-\dot{\phi}^2 r+v'(r)-j_r)=
e^{-i\int_{C^{(+)}}dt \hat{j}'_{r}\hat{e}_r} e^{i\int_{C^{(+)}}dt
j_{r} e_r} \prod_t \delta (\ddot{r}-\dot{\phi}^2 r+v'(r)-j'_{r})
\label{78} \end{equation}and \begin{equation}  \prod_t
\delta(\partial_t (\dot{\phi}r^2 )-rj_{\phi})= e^{-i\int_{C^{(+)}}dt
\hat{j}'_{\phi}\hat{e}_{\phi}} e^{i\int_{C^{(+)}}dt j_{\phi}r
e_{\phi}} \prod_t r(t)\delta(\partial_t (\dot{\phi}r^2 )-j'_{\phi}).
\label{79} \end{equation}Here $j_{r}$ and $j_{\phi}$ was defined in
(\ref{77}). In result, we get to the path integral formalism written
in terms of cylindrical coordinates. This is a very simplification
which will help to solve a lot of mechanical problems. One can note
that in result of mapping our problem reduced to the description of
quantum fluctuations of the surface of cylinder: $$ R(E)=2\pi
\int^{\infty}_{0}dT e^{\frac{1}{2i}\hat{\omega}\hat{\tau}-
i\int_{C^{(+)} (T)}dt ( \hat{j}_{r}(t)\hat{e}_{r}(t)+
\hat{j}_{\phi}(t)\hat{e}_{\phi}(t))}\t$$\begin{equation}  \t \int
D^{(2)}M(r,\phi) e^{-i\tilde{H}_T (x;\tau)-iU_T (x,e_{C})},
\label{80}
\end{equation}where \ba D^{(2)}M(r, \phi ) =\delta (E+\omega -H_T (r,\phi ))
\prod_t r^2 (t)dr(t)d\phi (t)\times \n \\ \times
 \delta (\ddot{r}-\dot{\phi}^2 r+v'(r)-j_{r})
\delta(\partial_t (\dot{\phi}r^2 )-j_{\phi}) \label{81} \ea and
\begin{equation} e_{C,1} =e_r \cos\phi -re_{\phi} \sin\phi,\;\;\;\;
e_{C,2}=e_r \sin\phi +re_{\phi} \cos\phi. \label{82} \end{equation}This is the
final result. The transformation looks quite classically but
(\ref{80}) can not be deduced from naive  coordinate transformation
of initial path integral for amplitude.

Inserting \begin{equation}  1=\int Dp Dl \prod_{t} \delta (p-\dot{r})
\delta (l-\dot{\phi}r^2) \label{83} \end{equation}into (\ref{80}) we
can introduce the motion in the phase space with Hamiltonian
\begin{equation} H_{j}= \frac{1}{2}p^2
+\frac{l^2}{2r^2}+v(r)-j_{r}r-j_{\phi}\phi. \label{84}
\end{equation}The Dirac's measure becomes four dimensional: $$
D^{(4)}M(r, \phi ,p,l) =\delta (E+\omega -H_T (r,\phi ,p,l)) \prod_t
dr(t)d\phi (t) dp(t) dl(t)\times $$\begin{equation} \times
\d\le(\dot{r}-\frac{\partial H_j}{\partial p}\ri) \d\le
(\dot{\phi}-\frac{\partial H_j}{\partial l}\ri) \d\le
(\dot{p}+\frac{\partial H_j}{\partial r}\ri) \d\le
(\dot{l}+\frac{\partial H_j}{\partial \phi}\ri) \label{85}
\end{equation}Note absence of the coefficient $r^2$ in this
expression. This is the result of special choice of transformation
(\ref{79}).

Since the Hamilton's group manifolds are more rich then Lagrange ones
the measure (\ref{85}) can be  considered as the starting point of
farther transformations. One must to note that the $(action,angle)$
variables are mostly useful \C{arnoldd}. Note also that to avoid the
technical problems with equations of motion and with functional
determinants it is useful to linearize the argument of $\d$-functions
in (\r{85}) hiding nonlinear terms in the corresponding auxiliary
variables $e_c$.

\subsection{Conclusions}

1. Our perturbation theory describes the quantum fluctuations of
the parameters $(h,\th)$ of classical trajectory $x_c$. It is more
complicated than canonical one, over an interaction constant
\C{tmph}, since demands investigation of analytic properties of
$4N$-dimensional integrals, where $2N$ is the phase space
dimension. Indeed, in the considered case with $N=1$ the
perturbations generating operator, $\kb$, see (\r{62}), contain
derivatives over four auxiliary parameters,
$(j_h,e_h,j_\th,e_\th)$.

Our transformed theory describes the "direct" deformations of
classical trajectory $x_c=x_c(h,\th)$, i.e. just $h$ and $\th$ are
the objects of quantization in the considered example. In another
words, the quantum deformations of the invariant hypersurface,
$(h,\th)$, is described in the new quantum theory. This
possibility is the consequence of $\d$-likeness of measure, i.e.
it based on the conservation of total probability.

Dirac measure allows to perform classical transformations of the
measure and to use high resources of classical mechanics. For
example, the interesting possibility may arise in connection with
Kolmogorov-Arnold-Mozer (KAM) theorem \C{KAM}: the system which is
not strictly integrable can show the stable motion peculiar to
integrable systems. This is the argument in favor of the idea that
there may be another, non-topological, mechanism of suppression of
the quantum excitations.

2. One can note that the transformed perturbation theory describes
only the retarded quantum fluctuations, see definition of Green
function (\r{66}). This feature of the theory can lead to the
imaginary time irreversibility of quantum processes and it must be
explained.

The starting expression (\r{31a}) describes the reversible in time
motion since total action $S_{C_+ (T_+ )}(x_+ ) - S_{C_- (T_- )}(x_-
)$ is time reversible. But the unitarity condition forced us to
consider the interference picture between expanding and converging
waves. This is fixed by the boundary conditions $e(0)=e(T)=0$. The
quantum theory remain time reversible up to canonical transformation
to the invariant hypersurface of the constant energy. The causal
Green function $G(t,t')$ , see (\ref{51b}), is able to describe both
advanced and retarded perturbations and the theory contains the
doubling of degrees of freedom. It means that the theory "keeps in
mind" the time reversibility. But after the canonical transformation,
using above mentioned boundary conditions, and continuing the theory
to the real time, the quantum perturbations were transferred on the
inner degrees of freedom of classical trajectory. In result the
memory of doubling of the degrees of freedom was disappeared and the
theory becomes "time irreversible".

The key step in this calculations was an extraction of the classical
trajectory $x_c$ which can not be defined without definition of
boundary conditions. Just $x_c$ introduces the direction of motion
and the order of quantum perturbations of trajectories inner degrees
of freedom play no role, i.e. the mechanical motion is time
reversible while the corrections to energy of trajectory, $h$, and to
the phase, $\th$, can not be time reversible. Therefore, the
considered irreversibility of the quantum mechanics in terms of
$(h,\th)$ seems to be imaginary.

\newpage

\section{Reduction of quantum degrees of freedom}\0

\subsection{Introduction}

It will be shown in this Section that the quantum fluctuations of
angular variables may be removed if the classical motion is periodic.
This cancelation mechanism can be used for path-integral explanation
of integrability of the quantum-mechanical problems, for example of
H-atom problem where the classical trajectories is closed
independently from the initial conditions\foot{The approach may be
extended on the case of rigid rotator problem \C {duru}. Last one is
isomorphic to the Pocshle-Teller problem \C{teler}}. The main result
of present Section is based on the statement that the topology
properties of classical trajectory takes special
significance\footnote{Since the action of perturbations generating
operator of transformedf theory, $\kb$, maps quantum corrections on
the boundaries of cotangent foliation, $\pa W$, see (\r{49c}).}.

Our technical problem consist in necessity to extract the quantum
angular degrees of freedom. For this purpose  we will define path
integral in the phase space of action-angle variables. For simplicity
the effect of cancelations we will demonstrate on the one-dimensional
$\lambda x^4$ model. In the following subsection the brief
description of unitary definition of the path-integral measure will
be given. The perturbation theory in terms of action-angle variables
will be contracted in Sec.4.3 (the scheme of transformed perturbation
theory was given firstly in \C{yadphys}). In Sec.4.4 the cancelation
mechanism will be demonstrated.

\subsection{Unitary definition of the path-integral measure}

We will calculate the probability \begin{equation}  R(E)= \int dx_1
dx_2 |A(x_1 ,x_2 ;E )|^2 , \label{1c} \end{equation}to introduce the
unitary definition of path-integral measure \C{yadphys}. Here
\begin{equation}  A(x_1 ,x_2 ;E )= i\int^{\infty}_{0} dT e^{iET}
\int_{x(0)=x_1}^{x(T)=x_2} Dx e^{iS_{C_+ (T)}(x)} \label{2}
\end{equation}is the amplitude of the particle with energy $E$ moving
from $x_1$ to $x_2$. The action \begin{equation}  S_{C_+ (T)}(x)=
\int_{C_+ (T)} dt (\frac{1}{2}\dot{x}^2 -
\frac{\omega_{0}^{2}}{2}x^2-\frac{\lambda}{4}x^4) \label{3}
\end{equation}is defined on the Mills' contour \C{millss}:
\begin{equation}  C_{\pm} (T): t\rightarrow t\pm i\epsilon,
\;\;\;\epsilon\rightarrow +0, \;\;\;0\leq t\leq T. \label{4}
\end{equation}So, we will omit the calculation of the amplitude.

Inserting (\ref{2}) into (\ref{1c}) we find, see previous Section,
that $$ R(E)=2\pi \int^{\infty}_{0} dT
e^{\frac{1}{2i}\hat{\omega}\hat{\tau}-i \int_{C^{(+)}(T)} dt
\hat{j}(t)\hat{e}(t)} \int Dx e^{-i\tilde{H}(x;\tau )-iU_T (x,e)}\t
$$\begin{equation} \t \d(E+ \omega -H_T (x)) \prod_{t} \d(\ddot{x}
+\o_{0}^{2}x +\la x^3 -j ). \label{11c} \end{equation}The "hat" symbol means
differentiation over corresponding auxiliary quantity. For instance,
\begin{equation}  \hat{\omega}\equiv\frac{\partial}{\partial \omega}, ~~~
\hat{j}(t)=\frac{\delta}{\delta j(t)}. \label{12z} \end{equation}It will be assumed
that $$ \hat{j}(t\in C_{\pm})j(t'\in C_{\pm})=\d(t-t'),
$$\begin{equation}  \hat{j}(t\in C_{\pm})j(t'\in C_{\mp})=0. \label{i}\end{equation}The time
integral over contour $C^{(\pm)}(T)$ means that \begin{equation}
\int_{C^{(\pm)}(T)} =\int_{C_{+}(T)} \pm \int_{C_{-}(T)}. \label{14}
\end{equation}At the end of calculations the limit $(\omega ,\tau, j, e)=0$
must be calculated. The explicit  form of $\tilde{H}(x;\tau )$
$U_T (x,e)$ will be given later; $H_T (x)$ is the Hamiltonian at
the  time moment $t=T$.

The functional $\delta$-function unambiguously determines the
contributions in the path integral. For this purpose we must find the
strict solution $x_j (t)$ of the equation of motion: \begin{equation}
\ddot{x} +\omega_{0}^{2}x +\lambda x^3 -j =0, \label{15}
\end{equation}expanding it over $j$. In zero order over $j$ we have
the classical trajectory $x_c$ which is defined by the equation of
motion: \begin{equation} \ddot{x} +\omega_{0}^{2}x +\lambda x^3  =0.
\label{16} \end{equation}This equation is equivalent to the following
one:
\begin{equation}  t+\theta_0 =\int^{x}dx \{2(h_0-\omega_{0}^{2}x^2
-\lambda x^4 ) \}^{-1/2}. \label{17} \end{equation}The solution of this equation is
the periodic elliptic function.

Here $(h_0, \theta_0)$ are the constants of integration of
Eq.(\ref{16}), i.e. $(h_0, \theta_0)$ are the coordinates of point on
the surface defined by elliptic function. The integration over $(h_0,
\theta_0)$ is assumed since the integration over all trajectories in
(\ref{2}) must be performed, i.e. $(h_0, \theta_0)$ takes on all
values available by elliptic function. Let $W$ be the corresponding
manyfold. One can say therefore that classical trajectory belongs $W$
completely.

The mapping of our problem on the action-angle phase space will be
performed using representation (\ref{11c}) \C{manj}. Using the
obvious definition of the  action: \begin{equation}
I=\frac{1}{2\pi}\oint \{2(h-\omega_{0}^{2}x^2 -\lambda x^4 )
\}^{1/2}, \label{18} \end{equation}and of the angle \begin{equation}
\phi =\frac{\partial h}{\partial I} \int^{x_c }
\{2(h-\omega_{0}^{2}x^2 -\lambda x^4 ) \}^{-1/2} \label{19}
\end{equation}variables \C{arnoldd} we easily find from (\ref{11c})
that $$ R(E)=2\pi \int^{\infty}_{0} dT
e^{\frac{1}{2i}\hat{\omega}\hat{\tau}-i \int_{C^{(+)}(T)} dt
\hat{j}(t)\hat{e}(t)}\int DI D\phi e^{-i\tilde{H}(x_c ;\tau )-iU_T
(x_c ,e)}\t$$\begin{equation} \t \delta (E+ \omega -h_T (I))
\prod_{t} \delta (\dot{I} -j\frac{\partial x_c}{\partial \phi})
\delta (\dot{\phi} -\Omega (I) +j\frac{\partial x_c}{\partial I}),
\label{20} \end{equation}where $x_c =x_c (I,\phi)$ is the solution of
Eq.(\ref{19}) with $h=h(I)$ as the solution of Eq.(\ref{18}) and the
frequency
\begin{equation} \Omega (I)=\frac{\partial h}{\partial I}. \label{21} \end{equation}Representation (\ref{20}) is not the full solution of our problem:
the action and angle variables are still interdependent since they
both are exited by the  same source $j(t)$. This reflects the
Lagrange nature of the path-integral description of phase-space
motion. The true Hamilton's description must contain independent
quantum sources of action and angle variables.

\subsection{Perturbation theory on the cotangent manifold}

The structure of source terms, $j\partial x_c/\partial \phi$ and
$j\partial x_c/\partial I$, show that the source of quantum
fluctuations is the classical trajectories perturbation and $j$ is
the auxiliary variable. It allows to regroup the perturbation series
in a following manner. Let us consider the action of the
perturbation-generating operators on $\d$-functions: $$
e^{-i\int_{C^{(+)}(T)} dt \hat{j}(t)\hat{e}(t) } e^{-iU_T (x,e)}
\prod_{t} \d\le(\dot{I} +j\frac{\partial x_c}{\partial \phi}\ri)
\d\le(\dot{\phi} -\Omega (I) -j\frac{\partial x_c}{\partial I}\ri)=
$$\begin{equation}  =\int D_{C^{(+)}}e_I D_{C^{(+)}}e_{\phi} e^{i\int_{C^{(+)}}dt
(e_I \dot{I}+e_{\phi}(\dot{\phi}-\Omega (I)))} e^{-iU_T (x,e_c )},
\label{22z} \end{equation}where \begin{equation}  e_c (e_I
,e_{\phi})= e_I \frac{\partial x_c}{\partial \phi} - e_{\phi}
\frac{\partial x_c}{\partial I}. \label{23c} \end{equation}The
integrals  over $(e_I ,e_{\phi})$ will be calculated perturbatively:
$$ e^{-iU_T (x,e_c )}= \sum^{\infty}_{n_I ,n_{\phi} =0} \frac{1}{n_I
!n_{\phi} !} \int \prod^{n_I }_{k=1}(dt_k e_I (t_k)) \prod^{n_{\phi}
}_{k=1}(dt'_k e_{\phi} (t'_k))\t$$\begin{equation} \t P_{n_I
,n_{\phi}} (x_c ,t_1 ,...,t_{n_I},t'_1,...,t_{n_{\phi}}), \label{24}
\end{equation}where
\begin{equation}  P_{n_I ,n_{\phi}} (x_c ,t_1
,...,t_{n_I},t'_1,...,t_{n_{\phi}})= \prod^{n_I }_{k=1} \hat{e}'_I
(t_k) \prod^{n_{\phi} }_{k=1} \hat{e}'_{\phi} (t'_k) e^{-iU_T (x,e'_c
)}, \label{25z} \end{equation}where $e'_c \equiv e_c (e'_I ,e'_{\phi}
)$ and the derivatives in (\r{25z}) are calculated at $e'_I =0$,
$e'_{\phi}=0$. At the same time, \begin{equation}  \prod^{n_I }_{k=1}
e_I (t_k) \prod^{n_{\phi} }_{k=1} e_{\phi} (t'_k)= \prod^{n_I }_{k=1}
(i\hat{j}_I (t_k)) \prod^{n_{\phi} }_{k=1} (i\hat{j}_{\phi} (t'_k))
e^{-i\int_{C^{(+)}} dt (j_I (t)e_I (t)+j_{\phi}(t)e_{\phi}(t))}.
\label{26z} \end{equation}The limit $(j_I , j_{\phi})=0$ is assumed.
Inserting (\ref{25z}), (\ref{26z}) into (\ref{22z}) we will find new
representation for $R(E)$: $$ R(E)=2\pi \int^{\infty}_{0} dT
e^{\frac{1}{2i}\hat{\omega}\hat{\tau}-i \int_{C^{(+)}(T)} dt
(\hat{j}_I (t)\hat{e}_I (t)+ \hat{j}_{\phi} (t)\hat{e}_{\phi}
(t))}\t$$$$\t \int DI D\phi e^{-i\tilde{H}(x_c ;\tau )-iU_T (x_c ,e_c
)}\t$$\begin{equation} \t \delta (E+ \omega -h_T (I)) \prod_{t}
\delta (\dot{I} -j_I ) \delta (\dot{\phi} -\Omega (I) -j_{\phi}),
\label{27c} \end{equation}in which the action and the angle are the
decoupled degrees of freedom.

Solving the canonical equations of motion: \begin{equation}  \dot{I}
=j_I,\;\;\;\dot{\phi} =\Omega (I) +j_{\phi} \label{28c}
\end{equation}the boundary conditions: \begin{equation}  I_j
(0)=I_0,\;\;\;\phi_j (0)=\phi_0 \label{29} \end{equation}will be
used. This will lead to the following Green function:
\begin{equation} g(t-t')=\Theta (t-t'), \label{30} \end{equation}with
boundary condition: $\Theta (0)=1$. The solutions of
eqs.(\ref{28c})  have the form: \ba I_j (t)=I_0 + \int dt' g(t-t')
j_I (t') \equiv I_0 +I'(t),
\n \\
\phi_j (t)=\phi_0 + \tilde{\Omega}(I_j) t+ \int dt' g(t-t') j_{\phi}
(t') \equiv \phi_0 +\tilde{\Omega}(I_0+I')t+\phi' (t), \label{31c}
\ea where \begin{equation}  \tilde{\Omega}(I_j ) =\frac{1}{t}\int dt'
g(t-t') \Omega (I_0 +I'(t')). \label{32z} \end{equation}Inserting
(\ref{31c}) into (\ref{27c}) we find: $$ R(E)=2\pi \int^{\infty}_{0}
dT e^{\frac{1}{2i}\hat{\omega}\hat{\tau}-i \int_{C^{(+)}(T)} dt
(\hat{j}{_I}(t)\hat{e}_I (t)+ \hat{j}_{\phi}(t)\hat{e}_{\phi}
(t))}\t$$\begin{equation} \t \int^{\infty}_{0} dI_0 \int^{2\pi}_{0}
d\phi_0 e^{-i\tilde{H}(x_c ;\tau )-iU_T (x_c ,e_c )} \delta (E+
\omega -h_T (I_j )), \label{33} \end{equation}where \begin{equation}
x_c =x_c (I_j ,\phi_j )= x_c (I_0 +I(t) ,\phi_0
+\tilde{\Omega}(I_0+I)t+\phi(t) ) \label{34c} \end{equation}and $e_c$
was defined in (\ref{23c}). Note that the measure of the integrals
over $(I_0 ,\phi_0 )$ was  defined without of the Faddeev-Popov's
ansatz and there is not any ``hosts" since the Jacobian of
transformation is equal to one.

We can extract the Green function into the perturbation-generating
operator using the equalities: \begin{equation}  \hat{j}_I (t)=\int
dt' g(t-t') \hat{I}(t), \hat{j}_{\phi}=\int dt' g(t-t')
\hat{\phi}(t), \label{35z} \end{equation}which evidently follows from
(\ref{31c}). In result, $$ R(E)=2\pi \int^{\infty}_{0} dT
e^{\{\frac{1}{2i}\hat{\omega}\hat{\tau} -i\int_{C^{(+)}(T)} dtdt'
g(t'-t) (\hat{I}(t)\hat{e}_I (t')+ \hat{\phi}(t)\hat{e}_{\phi}
(t'))\}}\t$$\begin{equation} \t \int^{\infty}_{0} dI_0
\int^{2\pi}_{0} d\phi_0 e^{-i\tilde{H}(x_c ;\tau )-iU_T (x_c ,e_c
)}\delta (E+ \omega -h_T (I_0+I)), \label{36} \end{equation}where
$x_c$ was defined in (\ref{34c}).

We can define the formalism without doubling of the degrees of
freedom. One can use fact that the action of
perturbation-generating operators and the analytical continuation
to the real times are commuting operations. This  can be seen
easily using the definition (\ref{i}). In result the expression:
$$ R(E)=2\pi \int^{\infty}_{0} dT
e^{\{\frac{1}{2i}\hat{\omega}\hat{\tau}-i \int_0^T dtdt' \Theta
(t'-t) (\hat{I}(t)\hat{e}_I (t')+ \hat{\phi}(t)\hat{e}_{\phi}
(t'))\}}\t$$\begin{equation} \t \int^{\infty}_{0} dI_0
\int^{2\pi}_{0} d\phi_0 e^{-i\tilde{H}(x_c ;\tau )-iU_T (x_c ,e_c )}
\delta (E+ \omega -h_T (I_0+ I(T)), \label{37} \end{equation}where
\begin{equation} \tilde{H}_T (x_c ;\tau )=
2\sum^{\infty}_{n=1}\frac{\tau^{2n+1}}{(2n+1)!}
\frac{d^{2n}}{dT^{2n}}h(I_0 +I(T)) \label{38} \end{equation}and
\begin{equation} -U_T (x_c ,e_c )=S(x_c +e_c )-S(x_c -e_c )-2\int_0^T
dt e_c\f{\d S(x_c)}{\d x_c}\label{39} \end{equation}defines quantum
theory on the cotangent manifold $W$.

Now we can use the last $\delta$-function:
$$ R(E)=2\pi \int^{\infty}_{0} dT
e^{\{\frac{1}{2i}(\hat{\omega}\hat{\tau}+ \int_0^T dtdt' \Theta
(t'-t) (\hat{I}(t)\hat{e}_I (t')+ \hat{\phi}(t)\hat{e}_{\phi} (t'))
\}}\t$$\begin{equation} \t\int^{\infty}_{0} dI_0 \int^{2\pi}_{0}
\frac{d\phi_0}{\Omega (E+\omega)} e^{-i\tilde{H}(x_c ;\tau )-iU_T
(x_c ,e_c )}. \label{40} \end{equation}Here \begin{equation}  x_c
(t)=x_c (I_0(E+\omega )+I(t)-I(T), \phi_0 +\tilde{\Omega}t+\phi (t)).
\label{41c}
\end{equation}
Eq.(\ref{40}) contains unnecessary contributions: the  action of the
operator \begin{equation}  \int^{T}_{0}dt dt' \Theta (t-t') \hat{e}_I
(t)\hat{I}(t') \label{42z} \end{equation}on $\tilde{H}_T$, defined in
(\ref{38}), leads to the time integrals with zero integration range:
$$ \int^{T}_{0}dt \Theta (T-t) \Theta (t-T) =0.$$ Using this fact,
\ba R(E)=2\pi \int^{\infty}_{0} dT e^{\frac{1}{2i}\int_0^T dtdt'
\Theta (t'-t) (\hat{I}(t)\hat{e}_I (t')+ \hat{\phi}(t)\hat{e}_{\phi}
(t'))}\times \n \\\times \int^{\infty}_{0} dI_0 \int^{2\pi}_{0}
\frac{d\phi_0}{\Omega (E)} e^{-iU_T (x_c ,e_c )}, \label{44z} \ea
where
\begin{equation}  x_c (t)=x_c (I_0(E)+I(t)-I(T), \phi_0 +\tilde{\Omega}t+\phi (t)).
\label{45z} \end{equation}is the periodic function: $$ x_c
(I_0(E)+I(t)-I(T), (\phi_0 +2\pi ) +\tilde{\Omega}t+\phi
(t))=$$\begin{equation} = x_c (I_0(E)+I(t)-I(T), \phi_0
+\tilde{\Omega}t+\phi (t)). \label{ii} \end{equation}Now we can
consider the cancelation of angular perturbations.

\subsection{Cancelation of angular perturbations}

{1. \it Simples example}

Introducing the perturbation-generating operator into the integral
over $\phi_0$: \ba R(E)=2\pi \int^{\infty}_{0} dT  e^{\frac{1}{2i}
\int_0^T dtdt' \Theta (t'-t) \hat{I}(t)\hat{e}_I (t')}\times \n
\\\times \int^{\infty}_{0} dI_0 \int^{2\pi}_{0} \frac{d\phi_0}{\Omega
(E)} e^{\frac{1}{2i}\int_0^T dtdt' \Theta (t'-t)
\hat{\phi}(t)\hat{e}_{\phi} (t')} e^{-iU_T (x_c ,e_c )}, \label{46z}
\ea the mechanism of cancelations of the angular perturbations
becomes evident. One can formulate  the statement:

(i) if \begin{equation}  e^{\frac{1}{2i}\int_0^T dtdt' \Theta (t'-t)
\hat{\phi}(t)\hat{e}_{\phi} (t')} e^{-iU_T (x_c ,e_c )}= e^{-iU_T
(x_c ,e_c )}|_{e_{\phi}=\phi =0} + dF(\phi_0 )/d\phi_0 , \label{49c}
\end{equation}and

(ii) if \begin{equation}  F(\phi_0 +2\pi )=F(\phi_0 ), \label{50}
\end{equation}then:
$$ R(E)=2\pi\int^{2\pi}_{0} \frac{d\phi_0}{\Omega (E)}
\int^{\infty}_{0} dT dI_0  e^{\frac{1}{2i}\int_0^T dtdt' \Theta
(t'-t) (\hat{I}(t)\hat{e}_I (t')}$$\begin{equation} \t e^{S(x_c
+e\partial x_c /\partial \phi_0 )- S(x_c -e\partial x_c /\partial
\phi_0 )}, \label{51}\end{equation}i.e. we find the expression in
which the angular corrections was canceled. In this case the problem
becomes semiclassical over the angular degrees of freedom.

For the $(\lambda x^4)_1$-model \begin{equation}  S(x_c +e\partial
x_c /\partial \phi_0 )- S(x_c -e\partial x_c /\partial \phi_0 )= S_0
(x_c)- 2\lambda \int^{T}_{0} dt x_c (t) \{e\partial x_c /\partial
\phi_0 \}^3 , \label{53} \end{equation}where \C{yadphys}
\begin{equation} S_0 (x_c)= \oint_T dt \le(\frac{1}{2}\dot{x}_c^2 -
\frac{\omega_{0}^{2}}{2}x_c^2-\frac{\lambda}{4}x_c^4\ri) \label{54}
\end{equation}is the closed time-path action and \begin{equation}
x_c (t)=x_c (I_0(E)+I(t)-I(T), \phi_0 +\tilde{\Omega}t). \label{52}
\end{equation}Here $I(t)$ and $I(T)$ are the auxiliary variables.

The condition (\ref{50}) requires that the  classical trajectory
$x_c$ with all derivatives over $I_0$, $\phi_0$ is the periodic
function. In the considered case of $(\lambda x^4)_1$-model $x_c$
is periodic function with period $1/\Omega$, see (\ref{ii}).
Therefore, we can concentrate the attention on the condition
(\ref{49c}) only.

Expanding $F(\phi_0)$ over $\lambda$: \begin{equation}  F(\phi_0)=
\lambda F_1 (\phi_0)+ \lambda^2 F_2 (\phi_0)+... \label{55}
\end{equation}we find that $$ \frac{d}{d\phi_0}F_1 (\phi_0)=$$$$=
\int^{T}_{0}\prod^{3}_{k=1}dt'_k \hat{\phi}(t'_k
)\le(\le(-\frac{6}{(2i)^3}\ri) \int^{T}_{0}dt \prod^{3}_{k=1}\Theta
(t-t'_k ) x_c (t)(\partial x_c /\partial I_0 )^3 e^{iS_0 (x_c )}_k
\ri) =
$$\begin{equation}  =\int^{T}_{0} dt' \hat{\phi}(t')B_1 (\phi ), \label{56}\end{equation}where $$
B_1 (\phi )=\le\{-\frac{6}{(2i)^3}\int^{T}_{0}dt \Theta
(t-t')\ri.\t$$\begin{equation}  \t\le.\prod^{2}_{k=1}(\Theta (t-t'_k
) \hat{\phi}(t'_k)) x_c (t)(\partial x_c /\partial I_0 )^3 e^{iS_0
(x_c )}\ri\} \end{equation}This example shows that the sum over all
powers of $\lambda$ can be written in the form: \begin{equation}
\frac{d}{d\phi_0}F(\phi_0)= \int^{T}_{0} dt' \hat{\phi}(t')B(\phi ),
\label{57z} \end{equation}where, using the definition (\ref{41c}),
\begin{equation} B(\phi )=\int^{T}_{0}dt \tilde{B}(\phi_0 +\phi (t)).
\label{58} \end{equation}Therefore, \begin{equation}
\hat{\phi}(t')B(\phi )=\frac{d}{d\phi_0} \int^{T}_{0} dt \delta
(t-t') \tilde{B}(\phi_0 +\phi (t)) \label{59} \end{equation}coincides
with the total derivative over initial phase $\phi_0$, and
\begin{equation}  F(\phi_0)=\tilde{B}(\phi_0 +\phi (t))|_{\phi=0}. \label{60} \end{equation}This
result ends the prove of (\r{49c}). \vskip 0.5cm {2. \it General
case}

Now we will offer following important statement:

--- {\it each order of perturbation theory in the invariant
subspace can be represented as the sum of total derivative over
the subspace coordinate.}\\ This statement directly follows from
structure of perturbations generating operator $\kb$ and the
assumption (\r{66b}). It explains the statement, offered in
Preface.

Let us remind that integration with last $\d$-function gives the
result of action of operator $\kb$ written in the form:
\begin{equation} R(E)=2\pi\int_0^\infty dT
\int_0^{2\pi}\f{d\vp_0}{\O(E)}:e^{-iU_(x_c,\he/2i)}:,\label{4.55.1}\end{equation}where
the colons mean normal product, \begin{equation} \he=\hj_\vp\f{\pa
x_c}{\pa I}-\hj_I\f{\pa x_c}{\pa\vp},\label{4.55.2}\end{equation}and
by definition $U_T$ is the odd over $\he_c$ functional:
\begin{equation} U_T(x_c,
e_c)=2\int_0^T\sum_{n=1}(\he_c(t)/2i)^{2n+1}u_n(x_c),\label{4.55}\end{equation}where
$u_n$ is the function of only $x_c$ at the time $t$. Inserting
(\r{4.55.2}) one can write: \begin{equation}
:e^{-iU_(x_c,\he/2i)}:=\prod_{n=1}^\infty
\prod_{k=0}^{2n+1}:e^{-iU_{k,n}(j,x_c)}:,\label{4.58}\end{equation}where
\begin{equation} U_{k,n}(j,x_c)=\int_0^T dt
(\hj_\vp(t))^{2n-k+1}(\hj_I(t))^{k} b_{k,n}(x_c(t))\label{4.59}\end{equation}and
the explicit form of $b_{k,n}(x_c)$ is not important.

Using the evident definition: $$ \hj_X=\int_0^T dt'
\Th(t-t')\hat{X}(t'),~~X=\vp, I, $$ it is easy to find that $$
j_X(t_1)b_{k,n}(x_c(t_2))=\Th(t_1-t_2)\pa b_{k,n}(x_c(t_2))/\pa
X_0,$$ since $x_c=x_c(X+X_0)$, or shortly: \begin{equation}
j_1b_2=\Th_{12}
\pa_{X_0}b_2=\pa_{X_0}(\Th_{12}b_2)\label{4.60}\end{equation}since
the indexes $(k,n)$ are not important.

Let us start consideration from the first term with $k=0$. In this
case we describe only the angular fluctuations. Noting that
$\pa_{X_0}$ and $\hj$ commute we can consider the lowest order over
$\hj$. The typical term looks as follows (omitting the index $X_0$):
$$ \hj_1\hj_2\cdots \hj_m b_1b_2\cdots b_m.$$ It is sufficient to
show that this expression is the total derivative over $X_0$.

Case $m=1$. In this approximation we have, see (\r{4.60}):
\begin{equation} \hj_1b_1=\Th_{11}\pa_0b_1\neq0.\label{4.61}\end{equation}Here
(\r{66b}) was used.

Case $m=2$. This order is less trivial: \begin{equation}
\hj_1\hj_2b_1b_2=
\Th_{21}b_1^2b_2+b_1^1b_2^1+\Th_{12}b_1b_2^2,\label{4.62}\end{equation}where
\begin{equation} b_i^n\equiv \pa^n b_i.\label{4.63}\end{equation}At first glance
(\r{4.62}) is not the total derivative. But inserting
$$1=\Th_{12}+\Th_{21}$$ we can symmetrize it: $$
\hj_1\hj_2b_1b_2=\Th_{21}(b_1^2b_2+b_1^1
b_2^1)+\Th_{12}(b_1b_2^2+b_1^1b_2^1)=$$$$=\pa_0(\Th_{21}b_1^1b_2
+\Th_{12}b_1b_2^1)\equiv$$\begin{equation} \equiv\pa_0(b_1^1\to
b_2+b_2^1\to b_1)\label{4.64}\end{equation}since the explicit form of
the function is not important. Therefore, the second order term can
be also reduced to the total derivative. Notice that (\r{4.64}) shows
time reversibility.

Case $m=3$. In this order one can find that \begin{equation}
\hj_1\hj_2\hj_3b_1 b_2b_3=\pa_0\le\{\sum_{i\neq j\neq k=1}^3 (i^2\to
j\to k+i^1\to j^1\to k)\ri\} \label{4.65}\end{equation} The $m$-th
order contribution is also total derivative: $$ \hj_1\hj_2\cdots
\hj_m b_1b_2\cdots b_m=\pa_0\{\sum_{i_1\neq i_2\neq i_3\neq\cdots\neq
i_m=1}^m(i_1^m\to i_2\to i_3\to\cdots \to i_m +$$$$+i_1^{m-1}\to
i_2^1\to i_3\to\cdots \to i_m + i_1^{m-2}\to i_2^1\to
i_3^1\to\cdots\to i_m +\cdots$$\begin{equation} \cdots + i_1^1\to
i_2^1\to i_3^1\to\cdots\to i_{m-1}^1\to
i_m)\}\label{4.66}\end{equation} Let us consider now the case with
$k\neq0$. The typical term looks as follows: \begin{equation}
\hj_1^1\hj_2^1\cdots\hj_l^1\hj_{l+1}^2\hj_{l+2}^2
\cdots\hj_m^2b_1b_2\cdots b_m,~0<l<m,\label{4.67}\end{equation}where,
for instance
\begin{equation}  \hj_k^1\equiv \hj_I(t_k),~~\hj_k^2\equiv \hj_\vp(t_k) \label{4.68}\end{equation}and \begin{equation}  \hj_1^ib_2=\Th_{12}\pa_0^ib_2.\label{4.69}\end{equation}
Case $m=2,l=1$.In this case: $$
\hj_1^1\hj_2^2b_1b_2=\Th_{21}(b_2\pa_0^1\pa_0^2b_1+(\pa_0^2b_2)
(\pa_0^1\pa_0^2b_1))+\Th_{12}(b_1\pa_0^1\pa_0^2b_2+(\pa_0^2b_2)
(\pa_0^1\pa_0^2b_1))=$$\begin{equation}
=\pa_0^1(\Th_{21}b_2\pa_0^2b_1+\Th_{12}
b_1\pa_0^2b_2)+\pa_0^2(\Th_{21}b_2\pa_0^1b_1+\Th_{12}b_1\pa_0^1b_2).
\label{4.70}\end{equation}Therefore we have the total-derivative
structure yet. This property is conserved in arbitrary order over $m$
and $l$ since the time-ordered structure does not depends from upper
index of $\hj$, see (\r{4.69}).

One can conclude that the contribution are defined by topology
properties of classical trajectory $x_c$. We will see that this
important property of perturbation theory remains unchanged also
for field theories with symmetry.

\subsection{Conclusions}

1. It was shown that the real-time quantum problem can be
semiclassical over the part of the degrees of freedom and quantum
over another ones. Following to the result of this Section one may
introduce the (probably naive) interpretation of the quantum systems
integrability (we suppose that the classical system is integrable and
can be mapped on the compact hypersurface in the phase space
\C{arnoldd}): the quantum system is strictly integrable in result of
cancelation of all quantum degrees of freedom. The mechanism of
cancelation of the quantum corrections is varied from case to case.

For some problems (as the rigid rotator, or the Pocshle-Teller)
the cancelation of angular degrees of the freedom is enough since
they carry only the angular ones. In an another case (as in the
Coulomb problem, or in the one-dimensional models) the problem may
be partly integrable since the quantum fluctuations of action
degrees of freedom just survive. Theirs absence in the Coulomb
problem needs special discussion (one must take into account the
dynamical (hidden) symmetry of Coulomb problem \C{popovv}).

The transformation to the action-angle variables maps the
$N$-dimensional Lagrange problem on the $2N$-dimensional phase-space
torus. If the winding number on this hypertorus is a constant (i.e.
the topological charge is conserved) one can expect the same
cancelations. This is important for the field-theoretical problems
(for instance, for sin-Gordon model \C{takhtajan}).

2. In the classical mechanics following approximated method of
calculations is used \C{arnoldd}. The canonical equations of motion:
\begin{equation}  \dot{I}=a(I,\phi),~~\dot{\phi}=b(I,\phi) \label{i'} \end{equation}are changed
on the averaged equations: \begin{equation}
\dot{J}=\frac{1}{2\pi}\int^{2\pi}_{0} d\phi a(J,\phi),~~
\dot{\phi}=b(J,\phi), \label{iix} \end{equation}It is possible if the
oscillations can be extracted from the systematic evolution of the
degrees of freedoms.

In our case \begin{equation}  a(I,\phi)=j\partial x_c /\partial
\phi,~~ b(I,\phi)=\Omega (I)- j\partial x_c /\partial I. \label{iii}
\end{equation}Inserting this definitions into (\ref{iix}) we find
evidently wrong result since in this approximation the problem looks
like pure semiclassical for the case of periodic motion:
\begin{equation} \dot{J}=0,~~\dot{\phi}=\Omega (J). \label{iv} \end{equation}The
result of this Section was used here. This shows that the procedure
of extraction of the oscillations from the systematic evolution is
not trivial and this method should be used carefully in the quantum
theories. (This approximation of dynamics is "good" on the time
intervals $\sim 1/|a|$ \C{arnoldd}.)

\newpage

\section{Example: H-atom}\0

\subsection{Introduction}

The mapping \begin{equation}  J:T \rar W,
\label{11}\end{equation}where $T$ is the $2N$-dimensional phase space
and $W$ is a linear space solves the mechanical problem iff
\begin{equation}  J=\otimes^N_1 J_i, \label{12f}\end{equation}where $J_i$
are the first integrals in involution, see e.g. \C{arnoldd}\foot{The
formalism of reduction (\r{11}) in classical mechanics is described
also in \C{mars}}. The aim of this Section is to adopt this procedure
for H-atom.

The mapping (\r{11}) introduces integral $manifold$
$J_{\o}=J^{-1}(\o)$ in such a way that the $classical$ phase space
flaw belongs to $J_{\o}$ $completely$. We wish quantize the $J_{\o}$
manifold instead of flow in $T$ noting that the quantum trajectory
also should belong to $J_{\o}$ completely. This important conclusion
was demonstrated in previous Section by transformation of the
path-integral measure to the canonical variables $(\x,\q)$. New
perturbation theory is extremely simple since $W$ is the linear
space.

The "direct" mapping (\r{11}) used in \C{yaph} assumes that $J$ is
known.  But it seems inconvenient having in mind the general
problem of nonlinear waves quantization, when the number of
degrees of freedom $N=\infty$, or if the transformation is not
canonical. We will consider by this reason the "inverse" approach
assuming that just the classical flow is known. Then, since the
flow belongs to $J_{\o}$ completely \C{yaph}, we would be able to
find the quantum motion in $W$. It is the main technical result
illustrated in this Section.

The manifold $J_{\o}$ is invariant relatively to some subgroup
$G_{\o}$ \C{smale} in accordance to topological class of classical
flaw. This introduces the $J_{\o}$ classification and summation
over all (homotopic) classes should be performed. Note, the
classes are separated by the boundary bifurcation lines in $W$
\C{smale}. If the quantum perturbations switched on adiabatically
then the homotopic group should stay unbroken. It is the ordinary
statement for quantum mechanics, but, generally speaking, this is
not true for field theories.

We will calculate the bound state energies in the Coulomb
potential\footnote{ We will restrict ourselves by the plane problem.
Corresponding phase space $T=(p,l,r,\vp)$ is 4-dimensional.}. This
popular problem was considered by many authors, using various
methods, see e.g. \C{popovv}. The path-integral solution of this
problem was offered firstly in \C{duruu}.

The classical flaw of this problem can be parameterized by the
angular momentum $l$, corresponding angle $\vp$ and by the normalized
on total Hamiltonian Runge-Lentz vector length $n$. So, we will
consider the mapping ($p$ is the conjugate to $r$ radial momentum in
the cylindrical coordinates): \begin{equation}
J_{l,n}:(p,l,r,\vp)\rar(l,n,\vp) \label{15z}\end{equation}to
construct the perturbation theory in the $W=(l,n,\vp)$ space. I.e.
$W$ is not considered as the cotangent foliation on $T$.

The mapping (\r{15z}) assumes additional reduction of the
four-dimensional incident phase space up to three-dimensional linear
subspace\footnote{$W$ would not have the simplectic structure.
Actually in considered case $W=R+TW$, where $R$ is the zero-modes
space and $TW$ is the simplectic subspace.}. Just this reduction
phenomena leads to corresponding stability of $n$ concerning quantum
perturbations and will allow to solve our H-atom problem
completely\footnote{ In other words, we would demonstrate that the
hidden Bargman-Fock \C{popovv} $O(4)$ symmetry is stay unbroken
concerning quantum perturbations.}.

In Subsec. 5.2 we will show how the mapping (\r{15z}) can be
performed for path-integral differential measure. In Subsec. 5.3 the
consequence of reduction will be derived and in Subsec. 5.4 the
perturbation theory in the $W$ space will be analyzed. The
calculations are based on the formalism offered in previous Sections.

\subsection{Mapping}

We will calculate the integral \C{yaph}: \begin{equation}
\R(E)=\int^{\infty}_0 dTe^{-i{\kb}(j,e)}\int DM(p,l,r,\vp)
e^{-iU(r,e)}, \label{21x}\end{equation}where $\R(E)$ is the
$probability$ to find a particle with energy $E$, i.e. we should find
\C{manj} that normalized on the zero-modes volume \begin{equation}
\R(E)=\pi\sum_n \d (E-E_n), \label{22}\end{equation}where $E_n$ are
the bound states energies. For $H$-atom problem $E_n\leq 0$.  This
condition will define considered homotopy class.

Expansion over operator \begin{equation}  {\kb}(j,e)=\f{1}{2}\int^T_0
dt (\h{j}_r\h{e}_r + \h{j}_{\vp}\h{e}_{\vp}),~~~ \h{X}(t)\equiv \d/\d
X(t), \label{23}\end{equation}generates the perturbation series. It
will be seen that in our case we may omit the question of
perturbation theories convergence.

The differential measure $$ DM(p,l,r,\vp)=\d (E-H_0)\prod_t dr(t)
dp(t) dl(t) d\vp (t)\t$$\begin{equation} \d\le(\dot{r}-\f{\pa
H_j}{\pa p}\ri)\d\le (\dot{p}+\f{\pa H_j}{\pa r}\ri)
\d\le(\dot{\vp}-\f{\pa H_j}{\pa l}\ri) \d\le(\dot{l}+\f{\pa H_j}{\pa
\vp}\ri), \label{24x}\end{equation}with total Hamiltonian
($H_0=H_j|_{j=0}$)
\begin{equation} H_j=\f{1}{2}p^2 -\f{l^2}{2r^2}-\f{1}{r}-j_r r
-j_{\vp}\vp \label{25}\end{equation}allows perform arbitrary transformation of
variables because of its $\d$-likeness. Notice that $H_j$ contains
only the "Lagrange forces" $j_r$ and $j_\vp$.

The functional $$ U(r,e)=-s_0(r)+$$\begin{equation} +\int^T_0
dt\le[\f{1}{((r+e_r)^2+r^2e_{\vp}^2)^{1/2}}-
\f{1}{((r-e_r)^2+r^2e_{\vp}^2)^{1/2}}+2\f{e_r}{r}\ri]
\label{26}\end{equation}describes the interaction between various
quantum modes and $s_0 (r)$ defines the non-integrable phase factor
\C{manj}. The quantization of this factor determines the bound state
energy. Such factor will appear if the phase of amplitude can not be
fixed \foot{As, for instance, in the Aharonov-Bohm case.}.  Note that
the Hamiltonian (\r{25}) contains the energy of radial $j_r r$ and
angular $j_{\vp}\vp$ excitation independently.

Let us introduce the functional $$ \D=\int \prod_t d^2\x d^2\eta\t
$$\begin{equation} \t\d (r(t)-r_c(\x,\eta))\d (p(t)-p_c(\x,\eta)) \d
(l(t)-l_c(\x,\eta))\d (\vp(t)-\vp_c(\x,\eta)) \label{27}\end{equation}which is defined by
given functions $(r_c,p_c,\vp_c,l_c)(\x,\eta)$. If given functions
$(\x,\q)$ zeroes argument of $\d$-functions in (\r{27}) then it is
assumed that the functional determinant \ba \D_c=\int \prod_t
d^2\bar{\x} d^2\bar{\eta} \d\le(\f{\pa
r_c}{\pa\x}\cdot\bar{\x}+\f{\pa r_c}{\pa\eta}
\cdot\bar{\eta}\ri)\d\le(\f{\pa p_c}{\pa\x}\cdot\bar{\x}+ \f{\pa
p_c}{\pa\eta}\cdot\bar{\eta}\ri)\times
\n \\
\times \d\le(\f{\pa \vp_c}{\pa\x}\cdot\bar{\x}+\f{\pa \vp_c}{\pa\eta}
\cdot\bar{\eta}\ri)\d\le(\f{\pa l_c}{\pa\x}\cdot\bar{\x}+ \f{\pa
l_c}{\pa\eta}\cdot\bar{\eta}\ri)\neq 0. \label{28}\ea Note that this is
the condition only for $(r_c,p_c,\vp_c,l_c)(\x,\eta)$.

To perform the mapping we will insert \begin{equation}  1=\D/\D_c
\label{transf}\end{equation}into (\r{21x}) and integrate over $r(t)$,
$p(t)$, $\vp(t)$ and $l(t)$. In result we find the measure: \ba
DM(\x, \eta)=\f{1}{\D_c}\d (E-H_0)\prod_t d^2\x d^2\eta
\d\le(\dot{r_c}-\f{\pa H_j}{\pa p_c}\ri)\times \n \\ \times \d
\le(\dot{p_c}+\f{\pa H_j}{\pa r_c}\ri) \d\le(\dot{\vp_c}-\f{\pa
H_j}{\pa l_c}\ri) \d\le(\dot{l_c}+\f{\pa H_j}{\pa \vp_c}\ri),
\label{29x}\ea Note that the functions $(r_c,p_c,\vp_c,l_c)(\x,\eta)$
must obey only one condition (\r{28}).

A simple algebra gives: \ba DM(\x,
\eta)=\f{\d(E-H_0)}{\D_c}\prod_td^2\x d^2\eta \int \prod_t
d^2\bar{\x}d^2\bar{\eta} \n \\ \times
\d^2\le(\bar{\x}-\le(\dot{\x}-\f{\pa h_j}{\pa\eta}\ri)\ri)
\d^2\le(\bar{\eta}-\le(\dot{\eta}+\f{\pa h_j}{\pa \x}\ri)\ri) \n
\\ \times \d\le(\f{\pa r_c}{\pa\x}\cdot\bar{\x}+\f{\pa r_c}{\pa\eta}
\cdot\bar{\eta} + \{r_c,h_j\}-\f{\pa H_j}{\pa p_c}\ri) \n\\\times
\d\le(\f{\pa p_c}{\pa\x}\cdot\bar{\x}+\f{\pa p_c}{\pa\eta}
\cdot\bar{\eta} + \{p_c,h_j\}+\f{\pa H_j}{\pa r_c}\ri) \n \\
\times \d\le(\f{\pa \vp_c}{\pa\x}\cdot\bar{\x}+\f{\pa \vp_c}{\pa\eta}
\cdot\bar{\eta}+ \{\vp_c,h_j\}-\f{\pa H_j}{\pa l_c}\ri) \n \\\times
\d\le(\f{\pa l_c}{\pa\x}\cdot\bar{\x}+\f{\pa l_c}{\pa\eta}
\cdot\bar{\eta}+ \{l_c,h_j\}+\f{\pa H_j}{\pa \vp_c}\ri).
\label{210}\ea The Poisson notation:
$$
\{X,h_j\}=\f{\pa X}{\pa \x}\f{\pa h_j}{\pa \eta}-
\f{\pa X}{\pa \eta}\f{\pa h_j}{\pa \x}
$$
was introduced in (\r{210}).

Next, the "auxiliary" quantity $h_j$ have been introduced by
following equalities: \ba \{r_c,h_j\}-\f{\pa H_j}{\pa p_c}=0,~
\{p_c,h_j\}+\f{\pa H_j}{\pa r_c}=0,
\n \\
\{\vp_c,h_j\}-\f{\pa H_j}{\pa l_c}=0,~ \{l_c,h_j\}+\f{\pa H_j}{\pa
\vp_c}=0. \label{211}\ea Then the functional determinant $\D_c$ is
canceled and \begin{equation}  DM(\x, \eta)=\d(E-H_0)\prod_td^2\x
d^2\eta \d^2(\dot{\x}-\f{\pa h_j}{\pa\eta}) \d^2(\dot{\eta}+\f{\pa
h_j}{\pa \x}), \label{212}\end{equation}It is the desired result of
transformation of the measure for given generating functions
$(r_c,p_c,\vp_c,l_c)(\x,\eta)$. In this case the "Hamiltonian" $h_j
(\x,\eta)$ is defined by four equations (\r{211}).

But there is another possibility. Let us assume that \begin{equation}
h_j (\x, \eta)=H_j (r_c, p_c, \vp_c, l_c)
\label{213}\end{equation}and the functions
$(r_c,p_c,\vp_c,l_c)(\x,\eta)$ are unknown. Then eqs.(\r{211}) are
the equations for this functions. It is not hard to see that the
eqs.(\r{211}) simultaneously with equations fixed by $\d$-functions
in (\r{212}) are equivalent of incident equations if the equality
(\r{213}) is hold. Indeed, for example, \begin{equation} \dot
r_c=\f{\pa r_c}{\pa \x}\cdot\dot\x+\f{\pa r_c}{\pa
\q}\cdot\dot\q=\{r_c,h_j\}=\f{\pa H_j}{\pa
p_c},\label{}\end{equation}where (\r{212}) and (\r{211}) was used
successively.

So, incident dynamical problem was divided on two parts. First one
defines the trajectory in the $W$ space through eqs.(\r{211}). Second
one defines the dynamics, i.e. the time dependence, through the
equations fixed by $\d$-functions in the measure (\r{212}).

Therefore, we should consider $r_c,~ p_c,~ \vp_c,~ l_c$ as the
solutions in the $\x,~\eta$ parametrization. The desired
parametrization of classical orbits has the form (one can find it in
arbitrary textbook of classical mechanics): \begin{equation}
r_c=\f{\eta_1^2(\eta_1^2+\eta_2^2)^{1/2}}
{(\eta_1^2+\eta_2^2)^{1/2}+\eta_2\cos \x_1},~ p_c=\f{\eta_2\sin
\x_1}{\eta_1(\eta_1^2+\eta_2^2)^{1/2}},~ \vp_c=\x_1,~l_c=\eta_1,
\label{214}\end{equation} i.e. $r_c$ and $p_c$ are $\x_2$
independent. At the same time, \begin{equation}
h_j=\f{1}{2(\eta_1^2+\eta_2^2)^{1/2}} -j_r r_c -j_\vp \x_1 \equiv h
(\eta)-j_r r_c -j_\vp \x_1. \label{215}\end{equation} Noting that the
derivatives of $h_j$ over $\x_2$ are equal to zero\footnote{To have
the condition (\r{28}) we should assume that $\pa r_c/\pa \x_2 \sim
\epsilon\neq 0$. We put $\epsilon=0$ completing the transformation.}
we find that $$ DM(\x, \eta)=\d(E-h(T))\prod_td^2\x d^2\eta
\d\le(\dot{\x}_1-\o_1+j_r\f{r_c}{\pa\eta_1}\ri) $$\begin{equation}
\times \d\le(\dot{\x}_2-\o_2+j_r\f{r_c}{\pa\eta_2}\ri)
\d\le(\dot{\eta}_1-j_r\f{\pa r_c}{\pa \x_1} -j_\vp\ri)
\d(\dot{\eta}_2), \label{217}\end{equation}where \begin{equation}
\o_i=\pa h/\pa\eta_i \label{218}\end{equation}are the conserved in
classical limit $j_r=j_\vp =0$ "velocities" in the $W$ space.

\subsection{Reduction}

We see from (\r{217}) that the length of Runge-Lentz vector is not
perturbated by the quantum forces $j_r$ and $j_{\vp}$. To investigate
the consequence of this fact it is useful to project this forces on
the axis of $W$ space. This means splitting of $j_r,~j_{\vp}$ on
$j_\x,~j_\eta$.  The equality
$$
\prod_t\d\le(\dot{\x}_1-\o_1+j_r\f{r_c}{\pa\eta_1}\ri)=
e^{\f{1}{2i}\int^T_0 dt \h{j}_{\x_1}\h{e}_{\x_1}} e^{2i\int^T_0 dt
j_r e_{\x_1}\pa r_c/\pa \eta_1}
\prod_t\d(\dot{\x}_1-\o_1+j_{\x_1})
$$
becomes evident if the Fourier representation of $\d$-function is
used (see also \C{yaph}). The same transformation of arguments of
other $\d$-functions in (\r{217}) can be applied. Then, noting that
the last $\d$-function in (\r{217}) is source-free, we find the same
representation as (\r{21x}) with \begin{equation}  \kb(j,e)=\int^T_0
dt (\h{j}_{\x_1}\h{e}_{\x_1}+ \h{j}_{\x_2}\h{e}_{\x_2}+
\h{j}_{\eta_1}\h{e}_{\eta_1}), \label{31}\end{equation}where the
operators $\h{j}$ are defined by the equality: \begin{equation}
\h{j}_X (t)=\int^T_0 dt' \Theta(t- t')\h{X}(t')
\label{44}\end{equation}and $\Theta(t- t')$ is the Green function of
our perturbation theory \C{yaph}.

We should change also \begin{equation}  e_r\rar e_c=e_{\eta_1}\f{\pa
r_c}{\pa \x_1}- e_{\x_1}\f{\pa r_c}{\pa \eta_1}- e_{\x_2}\f{\pa
r_c}{\pa \eta_2},~~e_\vp\rar e_{\x_1} \label{32}\end{equation}in the
Eq.(\r{26}). The differential measure takes the simplest form: $$
DM(\x, \eta)=\d(E-h(T))\prod_td^2\x d^2\eta
\d(\dot{\x}_1-\o_1-j_{\x_1}) \d(\dot{\x}_2-\o_2-j_{\x_2})
$$\begin{equation} \t \d(\dot{\eta}_1-j_{\eta_1}) \d(\dot{\eta}_2).
\label{33x}\end{equation}
Note now that the $\x, \eta$ variables are contained in $r_c$ only:
$$
r_c= r_c (\x_1, \eta_1, \eta_2).
$$
This means that the action of the operator $\h{j}_{\x_2}$ gives
identical to zero contributions into perturbation theory series. And,
since $\h{e}_{\x_2}$ and $\h{j}_{\x_2}$ are conjugate operators, see
(\r{31}), we can put
$$
j_{\x_2}=e_{\x_2}=0.
$$
This conclusion ends the reduction: \begin{equation}
\kb(j,e)=\int^T_0 dt (\h{j}_{\x_1}\h{e}_{\x_1}+
\h{j}_{\eta_1}\h{e}_{\eta_1}), \label{34}\end{equation}\begin{equation}
e_c=e_{\eta_1}\f{\pa r_c}{\pa \x_1}-e_{\x_1}\f{\pa r_c}{\pa \eta_1}.
\label{34'}\end{equation}The measure has the form: \begin{equation}  DM(\x,
\eta)=\d(E-h(T))d\x_2(0) d\eta_2(0)\prod_td\x_1 d\eta_1
\d(\dot{\x}_1-\o_1-j_{\x_1}) \d(\dot{\eta}_1-j_{\eta_1}) \label{35}\end{equation}since $V=V(r_c, e_c, \x_1)$ is $\x_2$ independent and
$$\int \prod_t dX(t)\d(\dot{X})=\int dX(0).$$

\subsection{Perturbations}

One can see from (\r{35}) that the reduction can not solve the
H-atom problem completely: there are nontrivial corrections to the
orbital degrees of freedom $\x_1,\eta_1$. By this reason we should
consider the expansion over $\kb$.

Using last $\d$-functions in (\r{35}) we find, see also \C{yaph}
(normalizing $\R(E)$ on the integral over $\x_2(0)\eta_2(0)$):
\begin{equation} \R(E)=\int^\infty_0 dT e^{-i\kb(j,e)}\int dM
e^{-iU(r_c,e)}, \label{41}\end{equation}where \begin{equation}
dM=\f{d\x_1 d\eta_1}{\o_2(E)}. \label{42}\end{equation}The operator
$\kb(j,e)$ was defined in (\r{34}) and $$
U(r_c,e_c)=-s_0(r)+$$\begin{equation} +\int^T_0
dt[\f{1}{((r_c+e_c)^2+ r_c^2e_{\x_1}^2)^{1/2}}
-\f{1}{((r_c-e_c)^2+r_c^2e_{\x_1}^2)^{1/2}}+2\f{e_c}{r_c}]
\label{a}\end{equation}with $e_c,~e_{\x_1}$ was defined in (\r{34'}),
(\r{32}) and
\begin{equation} r_c(t)=r_c(\eta_1 +\eta(t), \bar{\eta}_2(E,T),
\x_1+\o_1(t) +\x(t)),~~E\equiv h(\eta_1 +\eta(T), \bar{\eta}_2),
\label{43}\end{equation}where $\bar{\eta}_2(E,T)$ is the solution of equation
$E=h$.

The integration range over $\x_1$ and $\eta_1$ is as follows:
\begin{equation} 0\leq \x_1 \leq 2\pi,~~-\infty \leq \eta_1 \leq
+\infty. \label{45x}\end{equation}First inequality defines the principal domain of
the angular variable $\vp$ and second ones take into account the
clockwise and anticlockwise motions of particle on the Kepler orbits.

We can write: \begin{equation}  \R(E)=\int^\infty_0 dT \int dM
:e^{-iV(r_c,\h{e})}: \label{47}\end{equation}since the operator
$\ln\kb$ is linear over $\h{e}_{\x_1}, \h{e}_{\eta_1}$.  The colons
means "normal product" with differential operators staying to the
left of functions and $U(r_c,\h{e})$ is the functional of operators:
\begin{equation} 2i\h{e}_c=\h{j}_{\eta_1}\f{\pa r_c}{\pa \x_1}-
\h{j}_{\x_1}\f{\pa r_c}{\pa \eta_1},~~2i\h{e}_{\x_1}=\h{j}_{\x_1}.
\label{48}\end{equation}Expanding $U(r_c, \h{e})$ over $\h{e}_c$ and
$\h{e}_{\eta_1}$ we find:
\begin{equation} U(r_c,\h{e})=-s_0(r_c) +2\sum_{n+m \geq
1}C_{n,m}\int^T_0 dt
{\h{e}_c^{2n+1}\h{e}_{\eta_1}^m}\f{1}{r_c^{2n+2}}, \label{46}\end{equation}where
$C_{n,m}$ are the numerical constants. We see that the interaction
part presents expansion over $1/r_c$ and, therefore, the expansion
over $U$ generates an expansion over $1/r_c$.

In result, see Sec.4.5,\begin{equation}  \R(E)=\int^\infty_0 dT \int
dM \{e^{is_0 (r_c)} + B_{\x_1}(\x_1, \eta_1) + B_{\eta_1}(\x_1,
\eta_1)\}. \label{49}\end{equation}The first term is the pure
semiclassical contribution and last ones are the quantum corrections.
The functionals $B$ are the total derivatives: \begin{equation}
B_{\x_1}(\x_1, \eta_1)=\f{\pa}{\pa \x_1}b_{\x_1}(\x_1, \eta_1),~~
B_{\eta_1}(\x_1, \eta_1)=\f{\pa}{\pa \eta_1}b_{\eta_1}(\x_1, \eta_1).
\label{410}\end{equation}This means that the mean value of quantum
corrections in the $\x_1$ direction are equal to zero:
\begin{equation} \int^{2\pi}_0 d\x_1 \f{\pa}{\pa \x_1}b_{\x_1}(\x_1,
\eta_1) =0 \label{411}\end{equation}since $r_c$ is the closed trajectory
independently from initial conditions, see (\r{214}).

In the $\eta_1$ direction the motion is classical: \begin{equation}
\int^{+\infty}_{-\infty} d\eta_1 \f{\pa}{\pa \eta_1} b_{\eta_1}(\x_1,
\eta_1)=0 \label{412}\end{equation}since (i) $b_{\eta_1}$ is the
series over $1/r_c^2$ and (ii) $r_c \rar \infty$ when $|\eta_1| \rar
\infty$. Therefore, \begin{equation}  \R(E)=\int^\infty_0 dT \int dM
e^{is_0 (r_c)}. \label{413}\end{equation}This is the desired result.

Noting that
$$
s_0 (r_c)= kS_1 (E),~~k=\pm 1, \pm 2,...,
$$
where $S_1 (E)$ is the action over one classical period $T_1$:
$$
\frac{\partial S_1 (E)}{\partial E}=T_1 (E),
$$
and using the identity \C{manj}:
$$
\sum^{+\infty}_{-\infty} e^{inS_1 (E)} =
2\pi \sum^{+\infty}_{-\infty}\d (S_1 (E) - 2\pi n),
$$
we find: \begin{equation}  \R(E)=\pi \O \sum_{n} \d (E + 1/2n^2)
\label{416}\end{equation}where $\O$ is the zero-modes volume.

\subsection{Conclusions}

The demonstrated above mechanism of reduction is universal: one can
introduce from the very beginning the arbitrary number of coordinates
$(\x,\eta)$. But later on the formalism automatically, through
dependence of classical trajectory on coordinates of $W$, will
extract the necessary set of variables $(\x,\eta)$. At the same time
$\dim(\x,\eta)=\dim W$ and the integrals over other ones will give
the volume
$$V_0=\int \prod d\x(0)d\eta(0),$$ see (\r{35}) where $\dim
V_0=2$.

Notice that appearance of the "0-dimensional" integral measure
$$d\x_2(0)d\eta_2(0)$$ in (\r{35}) reflects the hidden $O(4)$
symmetry of H-atom problem \C{popovv}. Therefore, following our
selection rule, we must consider in a first place the classical
trajectory which leads to the maximal value of $\dim V_0$, i.e. we
must consider the contributions with maximal number of zero modes.

\newpage

\section{Example: sin-Gordon model}\0

\subsection{Introduction}

First of all we will describe "canonical" transformation in the
path-integral formalism. The method of canonical transformations in
spite of its expected effectiveness is unpopular in quantum theories
since on this way exist the problem: it is necessary to find the
transformation from Lagrangian to Hamiltonian descriptions. This
transition in general is very difficult if $\vp(x)$ and
$\dot{\vp}(x)=p(x)$ are not the independent quantities \C{dira}. But
we may use following trick. We start from the simplest verse of the
canonical formalism introducing the "first-order"
description\footnote{In other words, we will still stay in the frame
of Lagrangian formalism.} and after transformation come to
independent canonically conjugate pares, $(\x,\eta)$, i.e. come to
Hamiltonian description. It is evident that in general the
transformation
$$\vp_c: (\vp,p)\to(\x,\q)$$ will not be canonical. The formalism of
present Section is the same as in the H-atom problem but there is
some distinction.

We will continue in this Section description of influence of the
phase-space structure on the result of quantum-mechanical
measurements started in previous Sections. Now we will calculate the
expectation value of the "order parameter" (mass-shell particles
production vertex) $\Ga (q;u)$ \C{physrep}:
$$
\R (q)=<|\Ga (q;u)|^2>_u,
$$
where $q$ is the mass-shell ($q^2 =m^2$) particles momentum and
$<>_u$ means averaging over the field $u(x,t)$. Just the procedure of
averaging would be the object of our interest considering the quantum
Hamiltonian system with symmetry $G$. By definition, $\R$ is the
$probability$ to find one mass-shell particle. Certainly, $\R (q)=0$
on the sourceless vacuum but, generally speaking, $\R(q)\neq 0$ in a
field with nonzero energy density.

Calculations will be illustrated by the integrable (1+1)-dimensional
model with non-polynomial Lagrangian \begin{equation}
L=\f{1}{2}(\pa_{\mu}u)^2 + \f{m_h^2}{\la^2}[\cos(\la u)-1],
\label{1.13}\end{equation}We will consider following formulation of
the problem. Formally nothing prevents to linearize partly our
problem considering the Lagrangian \begin{equation}
L=\f{1}{2}[(\pa_{\mu}u)^2 - m_h^2 u^2] + \f{m_h^2}{\la^2}[\cos(\la
u)-1 + \f{\la^2}{2} u^2] \equiv L_0(u)-v(u)
\label{1.14}\end{equation}to describe creation (and absorption) of
the mass $m_h$ particles. Then the last term in (\ref{1.14}),
\begin{equation} v(u)=-\f{m_h^2}{\la^2}[\cos(\la u)-1 + \f{\la^2}{2}
u^2], \label{inter}\end{equation}describes interactions. The
corresponding to this theory order parameter is \begin{equation}  \Ga
(q;u)=\int dx dt e^{iqx} (\pa^2 +m_h^2)u(x,t),~~~q^2 =m_h^2.
\label{or}\end{equation}It will be shown by explicit calculations
that
\begin{equation}  \R (q) =0 \label{1z}\end{equation}as the consequence of unbroken
$\tilde{sl}(2,C)$ Kac-Moody algebra on which the solitons of theory
(\r{1.13}) live\footnote{Trivialness of soliton $S$-matrix was shown
in \C{zamol}}, see e.g. \C{sol} and references cited
therein\footnote{ It may be useful at this point to compare our
approach with ordinary thermodynamics of ferromagnetic. The external
magnetic field is $\sim<{\mu}>$, where the {\it order parameter}
$<\mu>$ is the mean value of the spin, and the phase transition means
that $<\mu>\neq 0$, i.e. $<\mu>=0$ means that corresponding symmetry
stay unbroken. We will suppose that the mean value of $|\Ga
(q,u)|^2$, which is the function of {external fields} parameter $q$,
play the same role for field theories with symmetry, i.e. $<|\Ga
(q,u)|^2>_u=0$ means that corresponding symmetry stay unbroken.
Therefore in our approach only the "external" display of symmetry can
be described.}. The solution (\r{1z}) seems interesting since it can
be interpreted as the explicit demonstration of field $u(x,t)$
confinement. The main purpose of this paper is to investigate how the
solution (\r{1z}) appears.

We will be able to find exact equality (\r{1z}) since the model
(\r{1.13}) possess infinite number of integrals of motion. It is well
known that each integral of motion in involution allows to shrink a
number of phase space $\bar{\ga}$ variables on two units, see e.g.
\C{arnoldd}. Resulting phase space $\ga$ is called as the reduced
phase space \C{mars}. The summation over all reduced phase space
topological classes \C{smale} is assumed.

By this way the field-theoretical problem will reduced to the
quantum-mechanical one. We would consider $\eta$ as the "particles"
generalized momentum and would introduce $\x$ as the conjugate to
$\eta$ coordinate of soliton.  The $2N$-dimensional phase space
(cotangent manifold) $\ga_N$ with local coordinates $(\x,\eta)$ on it
has natural simplectic structure, and $DM(\ga_N)=D^N M(\x,\eta)$ in
practical calculations (see Subsec.6.2). The summation over $N$ is
assumed.

The quantum corrections to semiclassical approximation of
transformed theory are simply calculable since $\eta$ are
conserved in the classical limit.  This is the particularity of
solitons dynamics (solitons momenta is the conserved quantities).
One can consider the developed in this paper formalism as the
path-integral version of nonlinear waves (solitons in our case)
quantum theory (the canonical quantization of sin-Gordon model in
the soliton sector was described also in \C{korepinn}.)

In Subsec.6.3 we will demonstrate Eq.(\r{1z}). It will be shown that
this solution is consequence of the previously developed proposition
(we would justify it in Subsec.6.2) that the semiclassical
approximation is exact for sin-Gordon model \C{dash}. The
semiclassical approximation in the $\ga_N$ phase space will be
considered in Subsec.6.2.

We would not use the complicated algebra to show the reduction
procedure explicitly noting that all solutions of model (\r{1.13})
are known \C{takhtajan}.  Then, using the $\d$-likeness of measure
$DM(\tilde{\ga})$, we will find in Subsec.6.2 $DM(\ga_N)$ considering
the mapping as an ordinary transformation to useful variables\foot{We
will apply inverse reduction procedure. Let $G$ be a group of
canonical transformations acting on the simplectic manifold
$\tilde\ga$ and let $\bar{G}$ be the Lie algebra of $G$ with $G^*$
dual of it. Then the momentum \C{sur} mapping $J:~\tilde{\ga}\rar
G^*$ introduces the integrals of motion which reduces the
$\tilde{\ga}$ manifold.  Noting that the set of levels
$J^{-1}({\eta})$ (solution of equations $J(\pi) =\eta$, $\pi \in
\tilde\ga$) is a manifold then $\ga_{\eta} = J^{-1}({\eta})/
\bar{G}_{\eta}$ is the reduced phase space, where $\bar{G}_{\eta}$ is
the co-adjoint isotropy subgroup of $G$. Therefore, the differential
measure $dM=dM(\eta, \ga_{\eta})$ for reduced phase space. For
integrable mechanical systems (infinite dimensional as well, see e.g.
\C{takhtajan}) $\ga_{\eta}$ shrinks to the point and in this case
$dM=dM(\eta)$ is the measure of momentum manifold. Just this simplest
case would be considered working with Lagrangian (\r{1.13}) and more
general and interesting case with measure $DM=DM(\eta, \ga_{\eta})$,
$\ga_{\eta} \neq \emptyset$, will be considered later. So, the
reduction procedure of our Hamiltonian system with symmetry $G$ looks
like canonical transformation \C{sol}. This problem is nontrivial
since, generally speaking, $\dim\tilde \ga$ and $\dim\ga$ are not the
same for model (\r{1.13}).}. Corresponding perturbation theory, see
Subsec.6.3, in the momentum space $J$ was described in \C{yaph}. In
Subsec.6.2 the path-integral definition of $\R (q)$ will be given.

We would conclude (this is the main result) that a theory in the
"nonlinear waves" sector may be nontrivial ($\R \neq 0$) iff the
manifold  $\ga$ is not compact.

\subsection{Reduction procedure}

{\it 6.2.1. Introduction into formalism.}

Our aim is to calculate the integral: \begin{equation}  \R (q) =
e^{-i \kb (j,e)}\int DM (u,p) |\Ga (q;u)|^2 e^{iS_O (u)-i U(u,e)},
\label{2.1}\end{equation}where $\Ga (q;u)$ was defined in (\r{or}).
In this expression the expansion over operator \begin{equation}
\kb(j,e)=\Re\int_{C_+} dx dt \f{\d}{\d j(x,t)}\f{\d}{\d e(x,t)}
\equiv \Re\int_{C_+} dx dt \hj(x,t)\he(x,t)
\label{2.2z}\end{equation}generates the perturbation theory series.
We will assume that this series exist. The functionals $U(u,e)$ and
$S_O(u)$ are defined by the equalities: \ba V(u+e) - V(u-e)= U(u,e)+
\int dx dt e(x,t) v'(u),
\n \\
S_0(u+e)-S_0(u-e)=S_O(u)+\int dx dt e(x,t)(\pa^2+m_h^2)u(x,t).
\label{2.3}\ea The action $S_0(u)$ corresponds to the free part of
Lagrangian (\r{1.13}) and $V(u)$ describes interactions. The quantity
$S_O (u)$ is not equal to zero since the soliton configurations have
nontrivial topological charge (see also \C{yadphys}). All time
integrals in this expressions were defined on the Mills time contour
\C{millss}:
$$
2\Re\int_{C_+}=\int_{C_+} + \int_{C_-}
$$
and
$$
C_{\pm} :t \rar t \pm i\e,~~~\epsilon\rar +0,~~~-\infty \leq t \leq
+\infty,
$$
to avoid the possible light-cone singularities of the perturbation
theory. The variational derivatives in (\r{2.2z}) are defined by
the following way:
$$
\f{\d u(x,t\in C_i)}{\d u(x',t'\in C_j)}=\d_{ij}\d (x-x') \d (t-t'),~~~
i,j=+,-.
$$
The auxiliary variables $(j,e)$ must be taken equal to zero at the
very end of calculations.

Considering the first order formalism with new coordinates $(u,p)$
the measure $DM(u,p)$ has the form: \begin{equation}  DM(u,p)=
\prod_{x,t} du(x,t) dp(x,t) \d \le(\dot{u}-\frac{\d H_j (u,p)}{\d
p}\ri) \d \le(\dot{p}+\frac{\d H_j (u,p)}{\d u}\ri)
\label{2.4b}\end{equation}with the total "Hamiltonian"
\begin{equation} H_j(u,p)=\int dx \le\{ \frac{1}{2} p^2 +\frac{1}{2}
(\pa_x u)^2 - \f{m_h^2}{\la^2}[\cos(\la u)-1] -ju\ri\}.
\label{2.5}\end{equation}The problem will be considered assuming that
$u(x,t)$ belongs to Schwartz space:
\begin{equation} u(x,t)|_{|x| =\infty}=0~({\rm{mod}}
\frac{2\pi}{\la}). \label{2.7}\end{equation}This means that $u(x,t)$ tends to zero
$(\rm{mod} \frac{2\pi}{\la})$ at $|x| \rar \infty$ faster then any
power of $1/|x|$. Note that $\dot u=p$, i.e. $u$ and $p$ are not the
independent quantities.

The measure (\r{2.4b}) allows to perform arbitrary transformations.
But, as was explained in Introduction, we will use the analog of
canonical transformation which conserves the form of equations of
motion. Hence, it is sufficient on this stage of calculations to know
only the fact that this transformation exist \C{takhtajan}. One may
propose that in result we should find for $N$-soliton topology:
\begin{equation} D^N M (\x ,\eta)=\prod_{t}d^N \x (t) d^N \eta (t)
\d^{(N)}\le(\dot{\x}-\frac{\pa h_j (\x ,\eta )}{\pa \eta (t)}\ri)
\d^{(N)}\le(\dot{\eta}+\frac{\pa h_j (\x ,\eta )}{\pa \x (t)}\ri),
\label{2.8}\end{equation}where $h_j$ is the "transformed Hamiltonian":
\begin{equation}  h_j =h_N (\eta) -\int dx j (x,t) u_N (x;\x ,\eta )
\label{2.9}\end{equation}and $u_N (x;\x ,\eta )$ is the $N$-soliton
configuration the time dependence of which is parameterized by $(\x
,\eta )$. Therefore, the local coordinates $(\x , \eta)$ are defined
by the equations: \begin{equation}  \dot{\x} =\f{\pa h_j}{\pa
\eta},~~~ \dot{\eta} =-\f{\pa h_j}{\pa
\x},\label{2.24b}\end{equation}where $h_j$ must obey the Poisson
conditions\footnote{See previous Section}:
\begin{equation}  \{u_c(x,t),h_j\}=\f{\d H_j}{\d p_c(x,t)},~~~
\{p_c(x,t),h_j\}=-\f{\d H_j}{\d u_c(x,t)}.
\label{2.17'}\end{equation}One can see choosing \begin{equation}  h_j
(\x,\eta) = H_j(u_c,p_c)\label{2.19}\end{equation}that the initial
equations have been restored:  $$ \dot{u}_c=\f{\pa u_c}{\pa
\x}\dot{\x} +\f{\pa u}{\pa \eta}\dot{\eta}= \{u_c,h_j\}=\f{\d H_j}{\d
p_c}.$$  The same we will have for $\dot{p}_c$. Therefore $(u_c,p_c)$
are solutions of equations of motion (\r{2.24b}), if the equality
(\r{2.19}) is hold.

The field theory case in $(1+1)$-dimensional configuration space
needs additional explanations. First of all, the analog of (\r{27})
must be introduced: \begin{equation}  \D(u,p)=\int \prod_t d^N \x (t)
d^N \eta (t) \prod_{x,t}\d (u(x,t)-u_c(x;\x ,\eta)) \d
(p(x,t)-p_c(x;\x ,\eta))\label{2.22}\end{equation}if the $N$-soliton
configuration is considered. Notice that the one-dimensional
$\d$-functions are introduced in (\r{2.22}) and $u_c$, $p_c$ are the
functions of sets $(\x,\eta)$, $\dim(\x,\eta)=2N$. Introducing
(\r{2.22}) we make the attempt to "hide" the time dependence entirely
into the set of $independent$ variables $(\x,\eta)$.

Comparing (\r{2.4b}) and (\r{2.8}) one can note that $x$ dependence
disappeared and the transformed measure depends on the number
$N=1,2,...$ Therefore, occurs the reduction of the quantum degrees of
freedom since the power of the coordinate set is continuum and the
number of solitons $N$ is the countable set. This means that the
proposed transformation to coordinates of solitons will be
unavoidably singular.

Notice then that the $x$ dependence of $\D(u,p)$ remain unimportant
since last one always appear under the integrals over all $u(x,t)$
and $p(x,t)$. At the same time it is important that introduced in
previous Section $\D_c$ disappeared in the final result, if the
integral form of Poisson brackets (\r{2.17'}) are hold\footnote{ See
the transformation (\r{transf}), described in previous Section. For
more confidence one can introduce the appropriate cells in the $x$
space \C{takhtajan}.}.

One can try to propose also the local form of canonical
commutators (\r{2.17'}), if the definition (\r{2.19}) is hold.
Indeed, one can find inserting (\r{2.19}) into (\r{2.17'}) that:
\begin{equation}  \{u_c(x,t),H_j(u_c,p_c)\}=\f{\d H_j(u_c,p_c)}{\d p_c(x,t)},~~~
\{p_c(x,t),H_j(u_c,p_c)\}=-\f{\d H_j(u_c,p_c)}{\d
u_c(x,t)}.\label{6.17}\end{equation}This equalities must hold for arbitrary $j$.
Making use the definition:
$$H_j(x_c,p_c)=\int dy \tilde{H}_j(x_c,p_c),$$ where $\tilde{H}_j$ is
the Hamiltonian density, one can write from (\r{6.17}):
$$\int dy\{u_c(x;\x ,\eta),u_c(y;\x ,\eta)\}\f{\d\tilde H_j}
{\d u_c(y,t)}+$$$$+ \int dy(\{u_c(x;\x ,\eta),p_c(y;\x
,\eta)\}-\d(x-y))\f{\d\tilde H_j}{\d p_c(y,t)}=0$$ and $$\int
dy\{p_c(x;\x ,\eta),p_c(y;\x ,\eta)\}\f{\d\tilde H_j} {\d
p_c(y,t)}-$$$$- \int dy(\{u_c(x;\x ,\eta),p_c(y;\x
,\eta)\}-\d(x-y))\f{\d\tilde H_j}{\d u_c(y,t)}=0.$$ Then one can
propose the solutions of these equations: $$ \{u_c(x;\x
,\eta),u_c(y;\x ,\eta)\}= \{p_c(x;\x ,\eta),p_c(y;\x ,\eta)\}=0,
$$\begin{equation}  \{u_c(x;\x ,\eta),p_c(y;\x ,\eta)\}=\d (x-y). \label{2.23}\end{equation}But it is interesting that the local commutators (\r{2.23}) are
not satisfied\foot{That circumstances was mentioned firstly by
V.Voronyuk.}. One can see this inserting the soliton solution into
(\r{2.23}). On the other hand the integral form (\r{6.17}) is
satisfied. All this means that $u_c$ and $p_c$ are not the
completely independent variables. It must be stressed that the
local relations (\r{2.23}) are not the necessary conditions in our
formalism.

In our terms, the quantum force $j(x,t)$ excites the $(\x, \eta)$
manifold only, leaving the topology of classical trajectory
$(u,p)_c$ unchanged. We can use them immediately since the
complete set of canonical coordinates $(\x ,\eta)$ of sin-Gordon
model is known, see e.g. \C{takhtajan}.

{\it 6.2.3. Perturbation theory on the cotangent bundle.}

The classical Hamiltonian $h_j$ is the sum: \begin{equation}  h_j
(\eta)=\int dp \sigma (r) \sqrt{r^2+m_h^2} +\sum^{N}_{i=1}h(\eta_i),
\label{4.10}\end{equation}where $\sigma (r)$ is the continuous
spectrum and $h(\eta)$ is the soliton energy. Note absence of
interaction energy among solitons.

New degrees of freedom $(\x,\eta)(t)$ must obey the equations
(\r{2.24b}): $$ \dot{\x}_i= \O (\eta_i) -\int dx j(x,t) \f {\pa u_N
(x;\x,\eta)}{\pa \eta_i},~~~ \O (\eta) \equiv \f {\pa h(\eta)}{\pa
\eta}, $$\begin{equation}  \dot{\eta}_i=\int dx j(x,t) \f {\pa u_N
(\x, \eta)}{\pa \x_i}. \label{4.11}\end{equation}Hence the sources of
quantum perturbations are proportional to the time-local fluctuations
of soliton configurations
$$
\f {\pa u_N (x;\x,\eta)}{\pa \eta_i},~~~\f {\pa u_N (x;\x,
\eta)}{\pa \x_i}.
$$
One can split the Lagrange source onto "Hamiltonian" ones:
$$j(x,t) \rar (j_{\x}, j_{\eta}).$$ This gives weight
functional $U(u_N;e_{\x},e_{\eta})$ and operator
$\kb(e_{\x},e_{\eta};j_{\x},j_{\eta})$. In result: \ba \R(q)=
\sum_N e^{-i \hat{K}(e_{\x},e_{\eta};j_{\x},j_{\eta})} \int D^N M
(\x,\eta)e^{iS_O(u_N)}e^{-iU(u_N;e_{\x},e_{\eta})}\times
\n \\
\times|\Ga (q;u_N)|^2 \label{4.13b}\ea where, using vector notations,
\begin{equation}  \kb(e_{\x},e_{\eta};j_{\x},j_{\eta}) =\f{1}{2}\int dt
\{\hat{j}_{\x} (t)\cdot \hat{e}_{\x}(t)+ \hat{j}_{\eta}(t)\cdot
\hat{e}_{\eta}(t)\}. \end{equation}The measure takes the form:
\begin{equation}  D^N M (\x,\eta)=\prod^{N}_{i=1}\prod_{t}d\x_i (t)
d\eta_i (t) \d (\dot{\x}_i- \O (\eta_i) - j_{\x,i}(t)) \d
(\dot{\eta}_i - j_{\eta,i}(t)) \label{b}\end{equation}The effective
interaction potential \begin{equation}
U(u_N;e_{\x},e_{\eta})=-\f{2m^2}{\la^2}\int dx dt \sin\la u_N~ (\sin
\la e -\la e) \label{4.15z}\end{equation}with \begin{equation}
e(x,t)=e_{\x}(t) \cdot \f{\pa u_N (x;\x,\eta)}{\pa \eta (t)}-
e_{\eta}(t) \cdot \f{\pa u_N (x;\x, \eta)}{\pa \x (t)}.
\label{4.15'z}\end{equation} Performing the shifts: \ba \x_i (t) \rar
\x_i (t) + \int dt' g(t-t') j_{\x,i}(t') \equiv \x_i (t) +\x'_i (t),
\n \\
\eta_i (t) \rar \eta_i (t) + \int dt' g(t-t') j_{\eta ,i}(t') \equiv
\eta_i (t) +\eta'_i (t), \label{4.16}\ea we can move the Green
function $g(t-t')$ into the operator: \begin{equation}
\kb(e_{\x},e_{\eta};{\x}',{\eta'}) =\f{1}{2}\int dt dt' g
(t-t')\{\hat{\x}'(t')\cdot \hat{e}_{\x}(t)+ \hat{\eta}'(t')\cdot
\hat{e}_{\eta}(t)\}. \end{equation}Notice that the Green function
$g(t-t')$ of eqs.(\ref{4.11}) is again the step function:
\begin{equation} g(t-t')=\Th (t-t') \label{4.12}\end{equation}Its imaginary part is
equal to zero for real times and this allows to shift $C_{\pm}$ to
the real-time axis (see \C{yaph}).

In result: \begin{equation}  D^N
M(\x,\eta)=\prod^{N}_{i=1}\prod_{t}d\x_i(t) d\eta_i(t) \d
(\dot{\x}_i- \O (\eta+\eta'))\d (\dot{\eta}_i)
\label{c}\end{equation}with
\begin{equation} u_N=u_N (x;\x+\x',\eta+\eta'). \label{4.19}\end{equation}The
equations: \begin{equation}  \dot{\x}_i=\O (\eta_i+\eta'_i)
\end{equation}are trivially integrable. In quantum case $\eta'_i \neq
0$ this equation describes the motion on nonhomogeneous and
anisotropic manifold. So, the expansion over
$(\hat{\x'},~\hat{e}_{\x},~\hat{\eta}', ~\hat{e}_{\eta})$ generates
the local in time deformations of $\ga_N$ manifold,
$(\x,\eta)\in\ga_N$ completely. The weight of this deformations is
defined by $U(u_N;e_{\x},e_{\eta})$.

Using the definition:
$$
\int Dx \d (\dot{x})=\int dx(0)=\int dx_0
$$
functional integrals are reduced to the ordinary integrals over
initial data $(\x,\eta)_{0}$. This integrals define the zero modes
volume.

\subsection {Quantum corrections}

The proof of (\r{1z}) we would divide on two parts. First of all
we would consider the semiclassical approximation (Sec.6.3.1) and
in Sec.6.3.2. we will show that this approximation is exact.

{\it 6.1. Introduction and definitions.}

The $N$-soliton solution $u_N$ depends from $2N$ parameters. Half
of them $N$ can be considered as the position of solitons and
other $N$ as the solitons momentum. Generally at
$|t|\rightarrow\infty$ the $u_N$ solution decomposed on the single
solitons $u_s$ and on the double soliton bound states $u_b$
\C{takhtajan}:
$$
u_N(x,t)=\sum^{n_1}_{j=1}u_{s,j}(x,t)+\sum^{n_2}_{k=1}u_{b,k}(x,t)+
O(e^{-|t|})
$$
We will see later that main elements of our formalism are the one
soliton $u_s$ and two-soliton bound state $u_b$ configurations. Its
$(\x,\eta)$ parameterizations, confirmed to eqs.(\r{2.17'}), have the
form: \begin{equation}
u_s(x;\x,\eta)=-\f{4}{\la}\arctan\{\exp(m_hx\cosh\b\eta -\x)\},~~~ \b
=\f{\la^2}{8} \label{so} \end{equation}and \begin{equation}
u_b(x;\x,\eta)=-\f{4}{\la}\arctan\{\tan\f{\b\eta_2}{2} \f{m_hx\sinh
\f{\b\eta_1}{2}\cos \f{\b\eta_2}{2}-\x_2} {m_hx\cosh
\f{\b\eta_1}{2}\sin \f{\b\eta_2}{2} -\x_1}\}. \label{bo}
\end{equation} The $(\x,\eta)$ parametrization of solitons individual
energies $h(\eta)$ takes the form:
$$
h_s(\eta)=\f{m_h}{\b}\cosh \b\eta,~~~
h_b(\eta)=\f{2m_h}{\b}\cosh \f{\b\eta_1}{2}\sin\f{\b\eta_2}{2}\geq 0.
$$
The bound-states energy $h_b$ depends from $\eta_1$ and $\eta_2$.
First one defines inner motion of two bounded solitons and second
one the bound states center of mass motion. Correspondingly we
will call this parameters as the internal and external ones. Note
that the inner motion is periodic, see (\r{b}).

Performing last integration in (\r{4.13b}) with measure (\r{c}) we
find: \begin{equation}  \R(q)= \sum_N \int \prod^{N}_{i=1} \{d\x_0
d\eta_0\}_i e^{-i \kb}e^{iS_O(u_N)}e^{-iU(u_N;e_{\x},e_{\eta})} |\Ga
(q;u_N)|^2 \label{5.1b}\end{equation}where \begin{equation}  u_N=u_N
(\eta_0 +\eta',\x_0 + \O (t) +\x'). \label{5.2}\end{equation}and
\begin{equation} \O (t)=\int dt'  \Th (t-t') \O (\eta_0 +\eta'(t'))
\label{5.3}\end{equation}
In the semiclassical approximation $\x'=\eta'=0$ we have:
\begin{equation} u_N=u_N (x;\eta_0 ,\x_0 + \O (\eta_0)t). \label{5.7}\end{equation}Note now that if the surface term \begin{equation}  \int
\pa_{\mu}(e^{iqx}\pa^{\mu}u_N)=0 \label{5.8b}\end{equation}then
\begin{equation}  \int d^2x e^{iqx}(\pa^2 +m_h^2)u_N (x,t) =
-(q^2-m_h^2)\int d^2x e^{iqx}u_N (x,t) =0 \label{5.9}\end{equation}since $q^2$
belongs to mass shell by definition. The condition (\ref{5.8b}) is
satisfied since $u_N$ belong to Schwartz space (the periodic boundary
condition for $u(x,t)$ do not alter this conclusion). Therefore, in
the semiclassical approximation (\r{1z}) is hold.

Expending the operator exponent in (\ref{5.1b}) we will find the
expansion over $$ \R_{n,m}(q)=\f{(1/2i)^n}{n!} \f{(1/2i)^m}{m!}
\lim_{(\x',\eta', e_\x,e_\eta)=0} \sum_N \int d^N \x_0 d^N \eta_0
\t$$$$\t\int \prod^{n}_{i=1}\{ dt_i dt'_i \th (t_i-t'_i)
\hat{\x}'(t'_i) $$$$ \times\int \prod^{m}_{i=1}\{ dt_i dt'_i \th
(t_i-t'_i) \hat{\eta}'(t'_i)\} e^{iS_O(u_N)}|\Ga (q;u_N)|^2
$$\begin{equation} \t \{\prod^{n}_{i=1}\hat{e}_{\x} (t_i)
\prod^{m}_{j=1}\hat{e}_{\eta} (t_j)
e^{-iU(u_N;e_{\x},e_{\eta})}\}|_{e=0}, \label{5.11}\end{equation}where
$U(u_N;e_{\x},e_{\eta})$ was defined in (\ref{4.15z}),
(\r{4.15'z}). Notice that the action of operators $\hat{\x}'$,
$\hat{\eta}'$ create terms \begin{equation}  \int d^2x e^{iqx} \th (t-t') (\pa^2
+m^2)u_N (x,t) \neq 0 .\label{5.12}\end{equation}
{\it 6.2. Quantum corrections}

Now we will show that {\it The semiclassical approximation is
exact in the soliton sector of (\r{1.13}), (\r{2.7}) theory.}

The structure of the perturbation theory is readily seen in the
"normal- product" form: \begin{equation}  \R(q)=\sum_N \int
\prod^{N}_{i=1} \{d\x_0 d\eta_0\}_i
:e^{-iU(u_N;\hat{j}/2i)}e^{iS_O(u_N)}|\Ga (q;u_N)|^2:,
\label{6.1b}\end{equation}where \begin{equation}
\hat{j}=\hat{j}_{\x}\cdot\f{\pa u_N}{\pa \eta}- \hat{j}_{\eta}\cdot
\f{\pa u_N}{\pa \x}=\o \hat{j}_{X}\f{\pa u_N}{\pa X}
\label{6.2b}\end{equation}and
\begin{equation}  \hat{j}_{X}=\int dt' \Th(t-t')\hat{X}(t')
\label{6.3}\end{equation}with $2N$-dimensional vector $X=(\x ,\eta)$. In Eq.
(\r{6.2b}) $\o$ is the ordinary simplectic matrix.

The colons in (\r{6.1b}) mean that the operator $\hat{j}$ should
stay to the left of all functions. The structure (\r{6.2b}) shows
that each order over $\hat{j}_{X_i}$ is proportional at least to
the first order derivative of $u_N$ over conjugate to $X_i$
variable.

The expansion of (\r{6.1b}) over $\hat{j}_{X}$ can be written
\C{yaph} in the form of total derivatives (omitting the semiclassical
approximation): \begin{equation}  \R(q)=\sum_N \int \prod^{N}_{i=1}
\{d\x_0 d\eta_0\}_i\le\{\sum^{2n}_{i=1}\f{\pa}{\pa
X_{0i}}P_{X_i}(u_N) \ri\}, \label{6.4b}\end{equation}where
$P_{X_i}(u_N)$ is the infinite sum of "time-ordered" polynomials (see
\C{yaph}) over $u_N$ and its derivatives. The explicit form of
$P_{X_i}(u_N)$ is complicated since the interaction potential is
non-polynomial. But it is enough to know, see (\r{6.2b}), that
\begin{equation} P_{X_i}(u_N)\sim \o_{ij} \f{\pa u_N}{\pa X_{0j}}.
\label{6.5b}\end{equation} Therefore, \begin{equation}  \R(q)=0
\label{6.6b}\end{equation}since (i) each term in (\r{6.4b}) is the
total derivative, (ii) we have (\r{6.5b}) and (iii) $u_N$ belongs to
Schwartz space.

We can conclude that the equality (\r{6.6b}) is hold since
\begin{equation} \f{\pa u_N}{\pa X_{0}}=0 ~~at~~X_0 \in \pa W,
\label{6.7}\end{equation}where $\pa W$ is the boundary of $W$.

In our consideration we did not touch the continuous spectrum
contributions. In considered approach this contributions are
absent since they are realized on zero measure: theirs
contributions are $\sim \{volume~of~\ga_N\}^{-1}$.

\newpage
\section{Summary}

Let as summarize the general results of present and of the
previous sections.

1. The $m$- into $n$-particles transition (non-normalized)
$probability$ $R_{nm}$ would have on the Dirac measure the
following symmetrical form}: $$ \R_{nm}(p_1,...,p_n, q_1,...,q_m)=
<\prod^{m}_{k=1}|\Ga(q_k;u)|^2 \prod^{n}_{k=1}|\Ga(p_k;u)|^2>_u=
$$$$ =e^{-i\hat{K}(j,e)}\int DM(u)e^{iS_O(u)-iU(u,e)}
\prod^{m}_{k=1}|\Ga(q_k;u)|^2 \prod^{n}_{k=1}|\Ga(p_k;u)|^2\equiv
$$\begin{equation}  \equiv\hat{\cal O}(u)\prod^{m}_{k=1}|\Ga(q_k;u)|^2
\prod^{n}_{k=1}|\Ga(p_k;u)|^2. \label{1.1e}\end{equation}Here $p(q)$ are the
in(out)-going particle momenta. It should be underlined that this
representation is strict and is valid for arbitrary Lagrange
theory of arbitrary dimensions.

2. The operator $\h{\cal O}$ contains three element. The Dirac
measure $DM$, the functionals $S_O$, $U(x,e)$ and the operator
$\kb(j,e)$.

The expansion over the operator \begin{equation}  \kb
(j,e)=\f{1}{2}\Re\int_{C_+} dx dt \f{\d}{\d j(x,t)}\f{\d}{\d e(x,t)}
\equiv \f{1}{2}\Re\int_{C_+} dx dt \hat{j}(x,t)\hat{e}(x,t)
\label{2.2'}\end{equation}generates the perturbation series. We will
assume that this series exist (at least in Borel sense).

3. The functionals $U(u,e)$ and $S_O(u)$ are defined by the
equalities: \begin{equation}  S_O(u)=(S_0(u+e)-S_0(u-e))
+2\Re\int_{C_+}dx dte(x,t)(\pa^2+m^2)u(x,t),
\end{equation}\begin{equation} U(u,e)= V(u+e)-V(u-e)-2\Re\int_{C_+}
dx dte(x,t)v'(u), \label{2.3a}\end{equation}where $S_0(u)$ is the
free part of the Lagrangian and $V(u)$ describes interactions. The
quantity $S_O(u)$ is not equal to zero if $u$ have nontrivial
topological charge.

4. The measure $DM(u,p)$ has the Dirac form: \begin{equation}
DM(u,p)= \prod_{x,t} du(x,t) dp(x,t) \d \le(\dot{u}-\frac{\d H_j
(u,p)}{\d p}\ri) \d \le(\dot{p}+\frac{\d H_j (u,p)}{\d u}\ri)
\label{2.4a}\end{equation}with the total Hamiltonian \begin{equation}
H_j(u,p)=\int dx \{ \f{1}{2}p^2 +\f{1}{2}(\nabla u)^2 + v(u)-ju \}.
\label{2.5a}\end{equation}This last one includes the energy $ju$ of
quantum fluctuations.

5. Dirac measure contains following information:

{\bf a.} Only $strict$ solutions of equations \begin{equation}
\dot{u}-\frac{\d H_j (u,p)}{\d p}=0,~ \dot{p}+\frac{\d H_j (u,p)}{\d
u}=0 \label{equ}\end{equation}with $j=0$ should be taken into
account. This "rigidness" of the formalism means the absence of
pseudo-solutions (similar to multi-instanton, or multi-kink)
contribution.

{\bf b.} $\R_{nm}$ is described by the $sum$ of all solutions of
Eq.(\r{equ}), independently from their "nearness" in the functional
space;

{\bf c.} $\R_{nm}$ did not contain the interference terms from
various topologically nonequivalent contributions. This displays
the orthogonality of corresponding Hilbert spaces;

{\bf d.} The measure (\r{2.4a}) includes $j(x)$ as the external
adiabatic source.  Its fluctuation disturbs the solutions of
Eq.(\r{equ}) and {\it vice versa} since the measure (\r{2.4a}) is
strict;

{\bf e.} In the frame of the adiabatical condition, the field
disturbed by $j(x)$ belongs to the same manifold (topology class)
as the classical field defined by (\r{equ}) \C{yaph}.

{\bf f.} The Dirac measure is derived for $real-time$ processes
only, i.e. (\r{2.4a}) is not valid for tunnelling ones.  For this
reason, the above conclusions should be taken carefully.

{\bf g.} It can be shown that theory on the measure (\r{2.4a})
restores ordinary (canonical) perturbation theory.

6. The parameter $\Ga(q;u)$ plays the role of particle production
vertex. It is connected directly with $external$ particle energy,
momentum, spin, polarization, charge, etc., and is sensitive to the
symmetry properties of the interacting fields system. For the sake of
simplicity, $u(x)$ is the real scalar field. The generalization would
be evident.

As a consequence of (\r{2.4a}), $\Ga(q;u)$ is the function of the
external particle momentum $q$ and is a $linear$ functional of
$u(x)$: \begin{equation}  \Ga(q;u)=-\int dx e^{iqx} \f{\d S_0 (u)}{\d
u(x)}= \int dx e^{iqx}(\pa^2 +m^2)u(x) ,~~q^2=m^2,
\label{1.2e}\end{equation}for the mass $m$ field. This parameter
presents the momentum distribution of the interacting field $u(x)$ on
the remote hypersurface $\s_\infty$ if $u(x)$ is the regular
function. Notice, the operator $(\pa^2 +m^2)$ cancels the mass-shell
states of $u(x)$.

The construction (\r{1.2e}) means, because of the Klein-Gordon
operator and since the external states being mass-shell by
definition \C{pei}, the solution $\R_{nm}=0$ is possible for a
particular topology (compactness and analytic properties) of
$quantum$ field $u(x)$. So, $\Ga(q;u)$ carries the following
remarkable properties:

-- it directly defines the observables,

-- it is defined by the topology of $u(x)$,

-- it is the linear functional of the actions symmetry group
element $u(x)$.

If (\r{equ}) have nontrivial solution $u_c(x,t)$, then this "extended
objects" quantization problem arises. We solve it introducing
convenient dynamical variables \C{jmp}. Then the measure (\r{2.4a})
admits the transformation: \begin{equation}  u_c:~(u,p)\to
(\x,\eta)\in W=G/G_c. \label{o11}\end{equation}and the transformed
measure has the form: \begin{equation}  DM(u,p)= \prod_{x,t\it C}
d\x(t) d\eta(t) \d \le(\dot{\x}-\frac{\d h_j (\x,\eta)}{\d\eta}\ri)
\d \le(\dot{\eta}+\frac{\d h_j (\x,\eta)}{\d\x}\ri),
\label{o12}\end{equation}where $h_j (\x,\eta)=H_j(u_c,p_c)$ is the
transformed Hamiltonian.

It is evident that $(\x,\eta)$ are parameters of integration of
eqs.(\r{equ}) and they form the factor space $W=G/G_c$. As a result
of mapping of the perturbation generating operator $\kb$ on the
manifold $W$ the equations of motion became linearized:
\begin{equation} DM=\prod_t\d\le(\dot{\x}-\frac{\d
h(\eta)}{\d\eta}-j_\x\ri) \d\le(\dot{\eta}-j_\eta\ri).
\label{o21}\end{equation}If Feynman's $i\e$-prescription is adopted,
then the Green function of Eq.(\r{o21}) \begin{equation}
g(t-t')=\Th(t-t') \label{o22}\end{equation}with boundary property:
$$\Th(0)=1.$$

7. Expansion of $\exp\{\kb(j,e)\}$ gives the "strong coupling'
perturbation series. Its analysis shows that the action of the
integro-differential operator $\h{\cal O}$ leads to the following
representation: \begin{equation}  \R_{nm}(p,q)= \int_W \{
d\x(0)\cdot\f{\pa}{\pa\x(0)}\R^\x_{nm}(p,q)+
d\eta(0)\cdot\f{\pa}{\pa\eta(0)}\R^\eta_{nm}(p,q)\}.
\label{o20}\end{equation}This means that the contributions into
$R_{nm}(p,q)$ are accumulated strictly on the boundary, "bifurcation
manifold", $\pa W$, i.e. depends directly on the topology property of
$W$.

8. It was shown that the MP is absent in the frame of Lagrangian
(\ref{1.13}). For this purpose one should modify the sin-Gordon
Lagrangian adding for instance the term: \begin{equation}
\frac{1}{2}(\pa \Phi)^2-\frac{1}{2}M^2\Phi^2-\frac{c}{3}u\Phi^2
\end{equation}to describe collision of "external" field $\Phi$ on the
solitons. This model allows to introduce the nontrivial probabilities
$\R (q_1,q_2, ...)$ considering creation (and absorption) of the
field $\Phi$. Note that field $u(x)$ is still "confined" even with
this adding.

\newpage

\section{Conclusion}

The final goal of present approach is to construct the workable at
arbitrary distances, i.e. for arbitrary momenta of produced hadrons,
$S$-matrix formalism for theories with (hidden) symmetry. But this
aim remains unachieved in present paper. In subsequent papers more
realistic field models in $4d$ Minkowski space-time metric will be
described. But one should not consider the demonstrated examples of
Yang-Mills $S$-matrix as the definite proves since I am note sure
that the used $O(4)\t O(2)$ solution of Yang-Mills equation in the
Minkowski in the situation of general position guarantee the largest
contribution. Moreover, only the $SU(2)$ theory will be considered.
Unfortunately we can not find in the frame of t'Hooft ansatz
\C{tooftans} the solution for larger $SU(N)$ group \C{vadim}.

It will be to shown how one or another physical phenomena may be seen
in the field theory with symmetry. Namely,

--- {\it no plain waves production exist in theories with symmetry},
\\ i.e. for instance the gluons can not be seen in a free state since
simply the last ones are absent in quantum theory of the symmetry
manifolds, or, in other words, since the gluon states and the
"states" of the symmetry manifold belong to the orthogonal Hilbert
spaces. The quark fields will not be included in this simplest
example. But more realistic model with quarks shows that

--- {\it inclusion of matter can not change previous conclusion that
the gluons can not be created.}\\ In the other example we will
show how the

--- {\it binding potential may arise among quarks.}\\ Here the
situation of general position selection rule will be extremely
important: it will be used that the situation when $(q\bar q)$
potential is independent from the scale of Yang-Mills fields is
mostly probable.

The quantum field theory with constraints will obey following
important property:

--- {\it the perturbation theory of quantum systems with symmetry
may be free from any divergences,}\\
i.e. it $may$\footnote{One can not be sure that the approach is
universal, can be used, for instance, in quantum gravity case.} be
rightful at arbitrary distances, for VHM case as well. It is the
evident consequence of lessening of the number of dynamical degrees
of freedom because of symmetry constraints\footnote{ And it is
unnecessary to have in that case any new mechanism, such as the
supersymmetry for example, to achieve the field theory without
divergences. Possible scenario of such theory will be discussed
later.}.

Exist also the intriguing question of asymptotic freedom. The
point is that there is no running coupling constants in our strong
coupling perturbation theory without divergences. On the other
hand the asymptotic freedom is the experimental fact. We will show
how

--- {\it the effect of asymptotic freedom may arise}\\
in our quantum theory of the symmetry manifolds. The main question
here is to find the experimentally observable corrections to the
asymptotic freedom law.

In summary, the aim of future publications would be the question: is
the offered approach complete from physical point of view? It is
important since offered quantization scheme in the situation of
general position on Dirac measure must be true for arbitrary
distances, since it is free from arbitrary scale
parameters\footnote{That is why I hope that it may give the
predictions acceptable from physical point of view at arbitrary
distances.}.

\newpage
\vskip 1cm {\bf Acknowledgments} \vskip 0.5cm

{\footnotesize First of all I am thankful to Alexei Sisakian  for
fruitful conversations during the work upon the topology conserving
perturbation theories ideology. The offered text was arranged under
last, before his sudden death, proposition to put in order my
present-day understanding of the approach. I would like to note the
significant role of E.Levin and L.Lipatov in realization of discussed
formalism. I am grateful to V.Kadyshevski for interest to the
discussed in the paper questions. Various parts of the approach were
offered to auditory of many Institutes and Universities and I am
grateful for theirs interest and comments.}

\end{document}